\documentclass[aps,prd,twocolumns,eqsecnum,preprintnumbers,showpacs,amsmath,amssymb]
{revtex4}

\usepackage{graphicx}
\usepackage{dcolumn}
\usepackage{bm}

\begin{document}

\title{THE MASS GAP AND SOLUTION OF THE \\ GLUON CONFINEMENT PROBLEM IN QCD}

\author{V. Gogokhia}
\email[]{gogohia@rmki.kfki.hu}

\affiliation{HAS, CRIP, RMKI, Depart. Theor. Phys., Budapest 114,
P.O.B. 49, H-1525, Hungary}

\date{\today}
\begin{abstract}
We propose to realize a mass gap in QCD not imposing the
transversality condition on the full gluon self-energy, while
preserving the color gauge invariance condition for the full gluon
propagator. Since due to color confinement the gluon is not a
physical state, none of physical observables/processes in
low-energy QCD will be directly affected by such a temporary
violation of color gauge invariance/symmetry. In order to make the
existence of a mass gap perfectly clear the corresponding
subtraction procedure is introduced. All this allows one to
establish the general structure of the full gluon propagator in
the presence of a mass gap. It is mainly generated by the
nonlinear interaction of massless gluon modes. The physical
meaning of the mass gap is to be responsible for the large-scale
(low-energy/momentum), i.e., nonperturbative structure of the true
QCD vacuum. The direct nonlinear iteration solution of the
transcendental equation for the full gluon propagator in the
presence of a mass gap is present. We formulate a general method
how to restore the transversality of the full gluon propagator
relevant for the nonperturbative QCD. It is explicitly shown that
such a solution confines QCD. The exact and gauge-invariant
criterion of gluon confinement is derived. The gauge-invariant
quark confinement criterion is also formulated.

\end{abstract}



\maketitle

\section{Introduction}

Quantum Chromodynamics (QCD) \cite{1,2} is widely accepted as a
realistic quantum field gauge theory of strong interactions not
only at the fundamental (microscopic) quark-gluon level but at the
hadronic (macroscopic) level as well. This means that in principle
it should describe the properties of experimentally observed
hadrons in terms of experimentally never seen quarks and gluons,
i.e., to describe the hadronic word from first principles -- an
ultimate goal of any fundamental theory. But this is a formidable
task because of the color confinement phenomenon, the dynamical
mechanism of which is not yet understood, and therefore the
confinement problem remains unsolved up to the present days. It
prevents colored quarks and gluons to be experimentally detected
as physical ("in" and "out" asymptotic) states which are colorless
(i.e., color-singlets), by definition, so color confinement is
permanent and absolute \cite{1}.

Today there is no doubt left that color confinement and other
dynamical effects, such as spontaneous breakdown of chiral
symmetry, bound-state problems, etc., being essentially
nonperturbative (NP) effects, are closely related to the
large-scale (low-energy/momentum) structure of the true QCD ground
state and vice-versa (\cite{3,4} and references therein). The
perturbation theory (PT) methods in general fail to investigate
them. If QCD itself is a confining theory then a characteristic
scale has to exist. It should be directly responsible for the
above-mentioned structure of the true QCD vacuum in the same way
as $\Lambda_{QCD}$ is responsible for the nontrivial perturbative
dynamics there (scale violation, asymptotic freedom (AF)
\cite{1}).

However, the Lagrangian of QCD \cite{1,2} does not contain
explicitly any of the mass scale parameters which could have a
physical meaning even after the corresponding renormalization
program is performed. So the main goal of this paper is to show
how the characteristic scale (the mass gap, for simplicity)
responsible for the NP dynamics in the infrared (IR) region may
explicitly appear in QCD. This becomes an imperative especially
after Jaffe and Witten have formulated their theorem "Yang-Mills
Existence And Mass Gap" \cite{5}. We will show that the mass gap
is dynamically generated mainly due to the nonlinear (NL)
interaction of massless gluon modes. In order to realize it our
proposal is to temporary violate the $SU(3)$ color gauge
invariance/symmetry of QCD. Since due to color confinement the
gluon is not a physical state, none of physical
observables/processes in low-energy QCD will be directly affected
by this proposal.

As mentioned above, there is no place for the mass gap in the QCD
Lagrangian, so the only place when the mass gap may appear is the
corresponding system of dynamical equations of motion, the
so-called Schwinger-Dyson (SD) equations, which should be
complemented by the Slavnov-Taylor (ST) identities (\cite{1} and
references therein). The propagation of gluons is one of the main
dynamical effects in the true QCD vacuum. It is described by the
above-mentioned corresponding SD quantum equation of motion for
the full gluon propagator. The importance of this equation is due
to the fact that its solutions reflect the quantum-dynamical
structure of the true QCD ground state. The color gauge structure
of this equation is also one of the main subjects of our
investigation in order to find a way how to realize a mass gap in
QCD.

The ultraviolet (UV) asymptotic (when the gluon momentum squared
goes to infinity ($q^2 \rightarrow \infty$)) of all the possible
solutions of the gluon SD equation is determined by AF (the number
of solutions is not fixed, since the equation is highly NL one).
At the same time, their IR asymptotic (when the gluon momentum
goes to zero ($q^2 \rightarrow 0$)) is, in general, of two types:
smooth and singular. Only the solution of the color confinement
problem will decide which type of the formal solutions really
takes place. It is explicitly shown here that indeed the general
iteration solution for the full gluon propagator in the presence
of a mass gap after making the necessary subtractions of the PT
contributions leads to the confining potential in QCD. It is
always IR singular, and thus the gluons remain massless. No
approximations/truncations and no special gauge choice for the
regularized skeleton loop integrals, contributing to the full
gluon self-energy, are made. The derived criterion of gluon
confinement is exact and gauge-invariant.

\section{QED}

It is instructive to begin with a brief explanation why a mass gap
does not occur in quantum electrodynamics (QED). The photon SD
equation can be symbolically written down as follows:

\begin{equation}
D(q) = D^0(q) + D^0(q) \Pi(q) D(q),
\end{equation}
where we omit, for convenience, the dependence on the Dirac
indices, and $D^0(q)$ is the free photon propagator. $\Pi(q)$
describes the electron skeleton loop contribution to the photon
self-energy (the so-called vacuum polarization tensor).
Analytically it looks

\begin{equation}
\Pi(q) \equiv \Pi_{\mu\nu}(q) = - g^2 \int {i d^4 p \over (2
\pi)^4} Tr [\gamma_{\mu} S(p-q) \Gamma_{\nu}(p-q, q)S(p)],
\end{equation}
where $S(p)$ and $\Gamma_{\mu}(p-q,q)$ represent the full electron
propagator and the full electron-photon vertex, respectively. Here
and everywhere below the signature is Euclidean, since it implies
$q_i \rightarrow 0$ when $q^2 \rightarrow 0$ and vice-versa. This
tensor has the dimensions of a mass squared, and therefore it is
quadratically divergent. To make the formal existence of a mass
gap (the quadratically divergent constant, so having the
dimensions of a mass squared) perfectly clear, let us now, for
simplicity, subtract its value at zero. One obtains

\begin{equation}
\Pi^s(q) \equiv \Pi^s_{\mu\nu}(q) = \Pi_{\mu\nu}(q) -
\Pi_{\mu\nu}(0) = \Pi_{\mu\nu}(q) - \delta_{\mu\nu}\Delta^2
(\lambda).
\end{equation}
The explicit dependence on the dimensionless UV regulating
parameter $\lambda$ has been introduced into the mass gap
$\Delta^2(\lambda)$, given by the integral (2.2) at $q^2=0$, in
order to assign a mathematical meaning to it. In this connection a
few remarks are in order in advance.
 The dependence on $\lambda$ (when it is not shown
explicitly) is assumed in all divergent integrals here and below
in the case of the gluon self-energy as well (see next section).
This means that all the expressions are regularized (including
photon/gluon propagator), and we can operate with them as with
finite quantities. $\lambda$ should be removed on the final stage
only after performing the corresponding renormalization program
(see below). Whether the regulating parameter $\lambda$ has been
introduced in a gauge-invariant way (though this always can be
achieved) or not, and how it should be removed is not important
for the problem if a mass gap can be "released/liberated" from the
corresponding vacuum. We will show in the most general way (not
using the PT and no special gauge choice will be made) that this
is impossible in QED and might be possible in QCD.

The decomposition of the subtracted vacuum polarization tensor
into the independent tensor structures can be written as follows:

\begin{equation}
\Pi^s_{\mu\nu}(q) = T_{\mu\nu}(q) q^2 \Pi^s_1(q^2) + q_{\mu}
q_{\nu}(q) \Pi^s_2(q^2),
\end{equation}
where both invariant functions $\Pi^s_n(q^2)$ at $n=1,2$ are, by
definition, dimensionless and regular at small $q^2$, since
$\Pi^s(0) =0$; otherwise they remain arbitrary. From this relation
it follows that $\Pi^s(q) = O(q^2)$, i.e., it is always of the
order $q^2$. Also, here and everywhere below

\begin{equation}
T_{\mu\nu}(q)=\delta_{\mu\nu}-q_{\mu} q_{\nu} / q^2 =
\delta_{\mu\nu } - L_{\mu\nu}(q).
\end{equation}

Taking into account the subtraction (2.3), the photon SD equation
becomes

\begin{equation}
D(q) = D^0(q) + D^0(q) \Pi^s(q) D(q) + D^0(q) \Delta^2(\lambda)
D(q).
\end{equation}
Its subtracted part can be summed up into geometric series, so one
has

\begin{equation}
D(q) = \tilde{D}^0(q) + \tilde{D}^0(q) \Delta^2(\lambda) D(q),
\end{equation}
where the modified photon propagator is

\begin{equation}
\tilde{D}^0(q) = {D^0(q) \over  1 - \Pi^s(q) D^0(q)}= D^0(q) +
D^0(q) \Pi^s(q) D^0(q) - D^0(q)\Pi^s(q)D^0(q) \Pi^s(q)D^0(q) + ...
\ .
\end{equation}
Since $\Pi^s(q) = O(q^2)$ and $D^0(q) \sim (q^2)^{-1}$, the IR
singularity of the modified photon propagator is determined by the
IR singularity of the free photon propagator, i.e.,
$\tilde{D}^0(q) = O (D^0(q))$ with respect to the behavior at
small photon momentum.

The photon self-energy (2.2) in terms of the independent tensor
structures is

\begin{equation}
\Pi_{\mu\nu}(q) = T_{\mu\nu}(q) q^2 \Pi_1(q^2) + q_{\mu} q_{\nu}
\Pi_2(q^2),
\end{equation}
where again $\Pi_n(q^2)$ at $n=1,2$ are dimensionless functions
and remain arbitrary. Due to the transversality of the photon
self-energy

\begin{equation}
q_{\mu} \Pi_{\mu\nu}(q) =q_{\nu} \Pi_{\mu\nu}(q) =0,
\end{equation}
which comes from the current conservation condition in QED, one
has $\Pi_2(q^2)=0$, i.e., it should be purely transversal

\begin{equation}
\Pi_{\mu\nu}(q) = T_{\mu\nu}(q) q^2 \Pi_1(q^2).
\end{equation}
On the other hand, from the subtraction (2.3) and the
transversality condition (2.10) it follows that

\begin{equation}
\Pi^s_2(q^2)= - {\Delta^2(\lambda) \over q^2},
\end{equation}
which, however, is impossible since $\Pi^s_2(q^2)$ is a regular
function of $q^2$, by definition. So the mass gap should be
discarded, i.e., put formally to zero and, consequently,
$\Pi^s_2(q^2)$ as well, i.e.,

\begin{equation}
\Delta^2(\lambda)=0, \quad \Pi^s_2(q^2)=0.
\end{equation}
Thus the subtracted photon self-energy is also transversal, i.e.,
satisfies the transversality condition

\begin{equation}
q_{\mu} \Pi_{\mu\nu}(q) =q_{\mu} \Pi^s_{\mu\nu}(q) =0,
\end{equation}
and coincides with the photon self-energy (see Eq. (2.3) at the
zero mass gap). Moreover, this means that the photon self-energy
does not have a pole in its invariant function $\Pi_1(q^2)=
\Pi^s_1(q^2)$. As mentioned above, in obtaining these results
neither the PT has been used nor a special gauge has been chosen.
So there is no place for quadratically divergent constants in QED,
while logarithmic divergence still can be present in the invariant
function $\Pi_1(q^2) = \Pi^s_1(q^2)$. It is to be included into
the electric charge through the corresponding renormalization
program (for these detailed gauge-invariant derivations explicitly
done in lower order of the PT see Refs. \cite{2,6,7,8,9}).

In fact, the current conservation condition (2.10), i.e., the
transversality of the photon self-energy lowers the quadratic
divergence of the corresponding integral (2.2) to a logarithmic
one. That is the reason why in QED only logarithmic divergences
survive. Thus in QED there is no mass gap and the relevant photon
SD equation is shown in Eq. (2.8), simply identifying the full
photon propagator with its modified counterpart. In other words,
in QED we can replace $\Pi(q)$ by its subtracted counterpart
$\Pi^s(q)$ from the very beginning ($\Pi(q) \rightarrow
\Pi^s(q)$), totally discarding the quadratically divergent
constant $\Delta^2(\lambda)$ from all the equations and relations.
The current conservation condition for the photon self-energy
(2.10), i.e., its transversality, and for the full photon
propagator $q_{\mu}q_{\nu}D_{\mu\nu}(q) = i\xi$ (here and
everywhere below $\xi$ is the gauge-fixing parameter) are
consequences of gauge invariance. They should be maintained at
every stage of the calculations, since the photon is a physical
state. In other words, at all stages the current conservation
plays a crucial role in extracting physical information from the
$S$-matrix elements in QED, which are usually proportional to the
combination $j^{\mu}_1 (q)D_{\mu\nu}(q) j^{\nu}_2(q)$. The current
conservation condition $j^{\mu}_1 (q) q_{\mu} =
j^{\nu}_2(q)q_{\nu} =0$ implies that the unphysical (longitudinal)
component of the full photon propagator does not change the
physics of QED, i.e., only its physical (transversal) component is
important. In its turn this means that the transversality
condition imposed on the photon self-energy is also important,
because $\Pi_{\mu\nu}(q)$ itself is a correction to the amplitude
of the physical process, for example such as electron-electron
scattering.

\section{QCD}

In QCD the gluon is not a physical state due to color confinement.
Still, color gauge invariance should also be preserved, so the
color current conservation takes place in QCD as well. However, in
this theory it plays no role in the extraction of physical
information from the $S$-matrix elements for the corresponding
physical processes and quantities. In other words, not the
conserved color currents, but only their color-singlet
counterparts, which can even be partially conserved, contribute
directly to the $S$-matrix elements describing this or that
physical process/quantity. For example, such an important physical
QCD parameter as the pion decay constant is given by the following
$S$-matrix element: $<0|J^i_{5\mu}(0)|\pi^j(q)>= i q_{\mu} F_{\pi}
\delta^{ij}$, where $J^i_{5\mu}(0)$ is just the axial-vector
current, while $|\pi^j(q)>$ describes the pion bound-state
amplitude, and $i, j$ are flavor indices.

So in QCD there is no such physical amplitude to which the gluon
self-energy may directly contribute (for example,
quark-quark/antiquark scattering is not a physical process). The
lesson which comes from QED is that if one preserves the
transversality of the photon self-energy at every stage, then
there is no mass gap. Thus, in order to realize a mass gap in QCD,
our proposal is not to impose the transversality condition on the
gluon self-energy, but preserving the color gauge invariance
condition for the full gluon propagator. As mentioned above, no
QCD physics will be directly affected by this. So color gauge
symmetry will be violated at the initial stage (at the level of
the gluon self-energy) and will be restored at the final stage (at
the level of the full gluon propagator, see below).

\subsection{Gluon SD equation}

The gluon SD equation can be symbolically written down as follows
(for our purposes it is more convenient to consider the SD
equation for the full gluon propagator and not for its inverse):

\begin{equation}
D_{\mu\nu}(q) = D^0_{\mu\nu}(q) + D^0_{\mu\rho}(q) i
\Pi_{\rho\sigma}(q; D) D_{\sigma\nu}(q),
\end{equation}
where $D^0_{\mu\nu}(q)$ is the free gluon propagator.
$\Pi_{\rho\sigma}(q; D)$ is the gluon self-energy, and in general
it depends on the full gluon propagator due to the non-Abelian
character of QCD. Thus the gluon SD equation is highly NL, while
the photon SD equation (2.1) is a linear one. In what follows we
omit the color group indices, since for the gluon propagator (and
hence for its self-energy) they are reduced to the trivial
$\delta$-function, for example $D^{ab}_{\mu\nu}(q) =
D_{\mu\nu}(q)\delta^{ab}$. Also, for convenience, we introduce $i$
into the gluon SD equation (3.1).

The gluon self-energy $\Pi_{\rho\sigma}(q; D)$ is the sum of a few
terms, namely

\begin{equation}
 \Pi_{\rho\sigma}(q; D)= - \Pi^q_{\rho\sigma}(q) -
\Pi^{gh}_{\rho\sigma}(q) + \Pi_{\rho\sigma}^t(D) +
\Pi_{(1)\rho\sigma}(q; D) + \Pi_{(2)\rho\sigma}(q; D) +
\Pi'_{(2)\rho\sigma}(q; D),
\end{equation}
where $\Pi^q_{\rho\sigma}(q)$ describes the skeleton loop
contribution due to quark degrees of freedom (it is an analog of
the vacuum polarization tensor in QED, see Eq. (2.2)), while
$\Pi^{gh}_{\rho\sigma}(q)$ describes the skeleton loop
contribution due to ghost degrees of freedom. Both skeleton loop
integrals do not depend on the full gluon propagator $D$, so they
represent the linear contribution to the gluon self-energy.
$\Pi_{\rho\sigma}^t(D)$ represents the so-called constant skeleton
tadpole term. $\Pi_{(1)\rho\sigma}(q; D)$ represents the skeleton
loop contribution, which contains the triple gluon vertices only.
$\Pi_{(2)\rho\sigma}(q; D)$ and $\Pi'_{(2)\rho\sigma}(q; D)$
describe topologically independent skeleton two-loop
contributions, which combine the triple and quartic gluon
vertices. The last four terms explicitly contain the full gluon
propagators in different powers, that is why they form the NL part
of the gluon self-energy. The explicit expressions for the
corresponding skeleton loop integrals \cite{10} (in which the
corresponding symmetry coefficients can be included) are of no
importance here. Let us note that like in QED these skeleton loop
integrals are in general quadratically divergent, and therefore
they should be regularized (see remarks above and below).

\subsection{A temporary violation of color gauge
invariance/symmetry (TVCGI/S)}

As we already know from QED, the regularization of the gluon
self-energy can be started from the subtraction its value at the
zero point (see, however, remarks below). Thus, quite similarly to
the subtraction (2.3), in QCD one obtains

\begin{equation}
\Pi^s_{\rho\sigma}(q; D) = \Pi_{\rho\sigma}(q; D) -
\Pi_{\rho\sigma}(0; D) = \Pi_{\rho\sigma}(q; D) -
\delta_{\rho\sigma} \Delta^2 (\lambda; D).
\end{equation}
Let us remind once more that for our purpose, namely to
demonstrate a possible existence of a mass gap $\Delta^2 (\lambda;
D)$ in QCD, it is not important how $\lambda$ has been introduced
and  how it should be removed at the final stage. The mass gap
itself is mainly generated by the nonlinear interaction of
massless gluon modes, slightly corrected by the linear
contributions coming from the quark and ghost degrees of freedom,
namely

\begin{equation}
\Delta^2 (\lambda; D)= \Pi^t(D) + \sum_a \Pi^a(0; D) =
\Delta^2_t(D) + \sum_a \Delta^2_a(0; D).
\end{equation}
Here index "a" runs as follows: $a= -q, -gh, 1, 2, 2'$, and $-q, \
- gh$ mean that both terms enter the above-mentioned sum with
minus sign (where, obviously, the tensor indices are omitted). In
these relation all the divergent constants $\Pi^t(D)$ and
$\Pi^a(0; D)$, having the dimensions of a mass squared, are given
by the corresponding skeleton loop integrals at $q^2=0$. We can
say that we parameterize the sum of all quadratic divergences as
the mass gap, and regularize it by $\lambda$. {\bf Since we are
not going to impose the transversality condition on the full gluon
self-energy, these constants summed up into the mass gap squared
$\Delta^2(\lambda; D)$ cannot be discarded like in QED, and
therefore the mass gap (3.4) should be explicitly taken into
account in QCD.} It is worth emphasizing here, that no
truncations/approximations and no special gauge choice are made
for the regularized skeleton loop integrals, contributing to the
full gluon self-energy and hence to the mass gap (3.4).

The general decomposition of the subtracted gluon self-energy into
the independent tensor structures can be written in he form

\begin{equation}
\Pi^s_{\rho\sigma}(q; D) = T_{\rho\sigma}(q) q^2 \Pi^s_1(q^2; D) +
q_{\rho} q_{\sigma} \Pi^s_2(q^2; D),
\end{equation}
where both invariant functions $\Pi^s_1(q^2; D)$ and $\Pi^s_2(q^2;
D)$ are dimensionless and regular at small $q^2$, since, by
definition, $\Pi^s_{\rho\sigma}(0; D) =0$ at any $D$. Evidently,
the subtracted quantities are free of the quadratic divergences,
but logarithmic ones can be still present in them like in QED.

\subsection{General structure of the full gluon propagator}

Taking into account the subtraction (3.3), the initial gluon SD
equation (3.1) becomes

\begin{equation}
D_{\mu\nu}(q) = D^0_{\mu\nu}(q) + D^0_{\mu\rho}(q)i
\Pi^s_{\rho\sigma}(q; D)D_{\sigma\nu}(q) + D^0_{\mu\sigma}(q)i
\Delta^2(\lambda; D) D_{\sigma\nu}(q).
\end{equation}
Introducing further the auxiliary free gluon propagator (see
Appendix A), the gluon SD equation (3.6), on account of the
decomposition (3.5), can be further simplified to

\begin{equation}
D_{\mu\nu}(q) = D^0_{\mu\nu}(q) - T_{\mu\sigma}(q)
\Bigl[\Pi^s_1(q^2; D) + { \Delta^2(\lambda; D) \over q^2} \Bigr]
D_{\sigma\nu}(q),
\end{equation}
where from now on $D^0_{\mu\nu}(q)$ is the standard free gluon
propagator, i.e.,

\begin{equation}
D^0_{\mu\nu}(q) = i \left\{ T_{\mu\nu}(q) + \xi L_{\mu\nu}(q)
\right\} {1 \over q^2}.
\end{equation}

The color gauge invariance condition imposed on the full gluon
propagator

\begin{equation}
q_{\mu}q_{\nu}D_{\mu\nu}(q) = i \xi,
\end{equation}
implies that it should be as follows:

\begin{equation}
D_{\mu\nu}(q) = i \left\{ T_{\mu\nu}(q) d(q^2) + \xi L_{\mu\nu}(q)
\right\} {1 \over q^2}.
\end{equation}
Substituting Eq. (3.10) into the gluon SD equation (3.7), one
obtains

\begin{equation}
d(q^2) = {1 \over 1 + \Pi^s_1(q^2; D) + (\Delta^2(\lambda; D) /
q^2)},
\end{equation}
and it is easy to check that this "solution" satisfies the initial
gluon SD equation (3.6), on account of the decomposition (3.5).

The only price we have paid by violating color gauge invariance is
the gluon self-energy, while the full and free gluon propagators
always satisfy it. Let us emphasize that the expression for the
full gluon form factor (3.11) cannot be considered as the formal
solution for the full gluon propagator, since both the mass gap
$\Delta^2(\lambda; D)$ and the invariant function $\Pi^s_1(q^2;
D)$ depend on $D$ themselves. In the formal PT limit
$\Delta^2(\lambda; D)=0$ the gluon self-energy coincides with its
subtracted counterpart, see Eq. (3.3), so everything will be like
in QED.

In what follows we will replace $\Pi^s_1(q^2; D) \rightarrow
\Pi(q^2; D)$, i.e., omitting, for convenience, both superscript
"s" and subscript "1".

\section{Nonlinear iteration solution}

 In order to perform a formal iteration of the gluon SD equation
(3.7), it is much more convenient to address its "solution" for
the full gluon form factor (3.11). Let us rewrite it in the form
of the corresponding transcendental (i.e., not algebraic)
equation, namely

\begin{equation}
d(q^2) = 1 - \Bigl[ \Pi(q^2; d) + {\Delta^2(\lambda; d) \over q^2}
\Bigr] d(q^2) = 1 - P(q^2; d) d(q^2),
\end{equation}
suitable for the formal nonlinear iteration procedure. Here we
replace the dependence on $D$ by the equivalent dependence on $d$.
For future purposes, it is convenient to introduce short-hand
notations as follows:

\begin{eqnarray}
\Delta^2(\lambda; d=d^{(0)} + d^{(1)} + d^{(2)} + ... + d^{(m)}+
...
) &=& \Delta^2_m = \Delta^2 c_m(\lambda, \alpha, \xi, g^2), \nonumber\\
\Pi(q^2; d=d^{(0)} + d^{(1)}+d^{(2)}+ ... + d^{(m)} + ...) &=&
\Pi_m(q^2),
\end{eqnarray}
and

\begin{equation}
P_m(q^2) = \Bigl[ \Pi_m(q^2) + {\Delta^2_m \over q^2} \Bigr], \
m=0,1,2,3,... \ .
\end{equation}
In these relations $\Delta^2_m$ are the auxiliary mass squared
parameters, while $\Delta^2$ is the mass gap itself. Via the
corresponding subscripts the dimensionless constants $c_m$ depend
on which iteration for the gluon form factor $d$ is actually done.
They may depend on the dimensionless coupling constant squared
$g^2$, as well as on the gauge-fixing parameter $\xi$. We also
introduce the explicit dependence on the dimensionless finite
(slightly different from zero) subtraction point $\alpha$, since
the initial subtraction at the zero point may be dangerous
\cite{1}. The dependence of $\Delta^2$ on all these parameters, as
well as on the number of different flavors $N_f$, is not shown
explicitly, and if necessary can be restored any time. Let us also
remind that all the invariant functions $\Pi_m(q^2)$ are regular
at small $q^2$. If it were possible to express the full gluon form
factor $d(q^2)$ in terms of these quantities then it would be the
formal solution for the full gluon propagator. In fact, this is
nothing but the skeleton loops expansion, since the regularized
skeleton loop integrals, contributing to the gluon self-energy,
have to be iterated. This is the so-called general iteration
solution. As mentioned above, no truncations/approximations and no
special gauge choice have been made. This formal expansion is not
a PT series. The magnitude of the coupling constant squared and
the dependence of the regularized skeleton loop integrals on it is
completely arbitrary.

It is instructive to describe the general iteration procedure in
some details. Evidently, $d^{(0)}=1$, and this corresponds to the
approximation of the full gluon propagator by its free
counterpart. Doing the first iteration in Eq. (4.1), one thus
obtains

\begin{equation}
d(q^2) = 1 - P_0(q^2) + ... = 1 + d^{(1)}(q^2) + ...,
\end{equation}
where obviously

\begin{equation}
d^{(1)}(q^2) = - P_0(q^2).
\end{equation}
Doing the second iteration, one gets

\begin{equation}
d(q^2) = 1 - P_1(q^2) [ 1 + d^{(1)}(q^2) ] + ... = 1 +
d^{(1)}(q^2) + d^{(2)}(q^2) + ...,
\end{equation}
where

\begin{equation}
d^{(2)}(q^2) = - d^{(1)}(q^2) - P_1(q^2) [ 1 - P_0(q^2)].
\end{equation}
Doing the third iteration, one further obtains

\begin{equation}
d(q^2) = 1 - P_2(q^2) [ 1 + d^{(1)}(q^2) + d^{(2)}(q^2)] + ... = 1
+ d^{(1)}(q^2) + d^{(2)}(q^2) + d^{(3)}(q^2) + ...,
\end{equation}
where

\begin{equation}
d^{(3)}(q^2) = - d^{(1)}(q^2) - d^{(2)}(q^2) - P_2(q^2) [ 1 -
P_1(q^2)(1 - P_0(q^2))],
\end{equation}
and so on for the next iterations.

Thus up to the third iteration, one finally arrives at

\begin{equation}
d(q^2) = \sum_{m=0}^{\infty} d^{(m)}(q^2) = 1 - [\Pi_2(q^2) +
{\Delta^2_2 \over q^2}] \Bigl[ 1 - [\Pi_1(q^2) + {\Delta^2_1 \over
q^2}] [1 - \Pi_0(q^2) - {\Delta^2_0 \over q^2}] \Bigr] + ... \ .
\end{equation}
We restrict ourselves by the iterated gluon form factor up to the
third term, since this already allows to show explicitly some
general features of the nonlinear iteration procedure.

\subsection{Splitting/shifting procedure}

Doing some tedious algebra, the previous expression can be
rewritten as follows:

\begin{eqnarray}
d(q^2) &=& [1 - \Pi_2(q^2) + \Pi_1(q^2) \Pi_2(q^2) - \Pi_0(q^2)
\Pi_1(q^2)\Pi_2(q^2) + ...] \nonumber\\
&+& {1 \over q^2} [\Pi_2(q^2)\Delta^2_1 + \Pi_1(q^2)\Delta^2_2 -
\Pi_0(q^2) \Pi_1(q^2)\Delta^2_2 - \Pi_0(q^2) \Pi_2(q^2)\Delta^2_1
- \Pi_1(q^2) \Pi_2(q^2)\Delta^2_2 + ...] \nonumber\\
&-& {1 \over q^4} [\Pi_0(q^2) \Delta^2_1 \Delta^2_2 + \Pi_1(q^2)
\Delta^2_0 \Delta^2_2 + \Pi_2(q^2) \Delta^2_0 \Delta^2_1 + ...]
\nonumber\\
&-& {1 \over q^2} [\Delta^2_2 -  {\Delta^2_1 \Delta^2_2 \over q^2}
+ { \Delta^2_0 \Delta^2_1 \Delta^2_2 \over q^4} + ...],
\end{eqnarray}
so that  this formal expansion contains three different types of
terms. The first type are the terms which contain only different
combinations of $\Pi_m(q^2)$ (they are not multiplied by inverse
powers of $q^2$); the third type of terms contains only different
combinations of $(\Delta^2_m / q^2)$. The second type of terms
contains the so-called mixed terms, containing the first and third
types of terms in different combinations. The two last types of
terms are multiplied by the corresponding powers of $1/q^2$.
Evidently, such structure of terms will be present in each
iteration term for the full gluon form factor. However, any of the
mixed terms can be split exactly into the first and third types of
terms by keeping the necessary number of terms in the Taylor
expansions in powers of $q^2$ for $\Pi_m(q^2)$, which are regular
functions at small $q^2$. Thus the IR structure of the full gluon
form factor (which just is our primary goal to establish) is
determined not only by the third type of terms. It gains
contributions from the mixed terms as well.

Let us present the above-mentioned Taylor expansions as follows:

\begin{equation}
\Pi_m(q^2) = \Pi_m(0) + (q^2 / \mu^2) \Pi^{(1)}_m (0) + (q^2 /
\mu^2)^2 \Pi^{(2)}_m (0) + O_m(q^6),
\end{equation}
since for the third iteration we need to use the Taylor expansions
maximum up to this order. Here and below $\mu^2$ is some fixed
mass squared (not to be confused with the tensor index), as well
as $O_m((q^2)^n), \ n=0,1,2,3,...$ denotes the functions which
Taylor expansions begin with the term of the order $(q^2)^n$;
otherwise they remain arbitrary. For example, for the mixed term
$(1/ q^2) \Pi_2(q^2)\Delta^2_1$ one has

\begin{equation}
{\Delta^2_1 \over q^2}\Pi_2(q^2) = {\Delta^2_1 \over q^2} \Bigl[
\Pi_2(0) + (q^2 / \mu^2) \Pi^{(1)}_2 (0) + O_2(q^4) \Bigr] =
{\Delta^2_1 \over q^2} \Pi_2(0) + a_1\Pi^{(1)}_2 (0) +O(q^2).
\end{equation}
Here and everywhere below $a_m = (\Delta^2_m / \mu^2), \
m=0,1,2,3,...$ are the dimensionless constants. The first term now
is to be shifted to the third type of terms and combined with the
term $(-1/q^2)\Delta^2_2$, while the second term $a_1\Pi^{(1)}_2
(0) +O(q^2)$ is to be shifted to the first type of terms. All
other mixed terms of similar structure should be treated
absolutely in the same way. For the mixed term $(-1 / q^4)
\Pi_0(q^2) \Delta^2_1 \Delta^2_2$, one has

\begin{eqnarray}
- {\Delta^2_1 \Delta^2_2 \over q^4}\Pi_0(q^2) &=& -{\Delta^2_1
\Delta^2_2 \over q^4} \Bigl[ \Pi_0(0) + (q^2 / \mu^2) \Pi^{(1)}_0
(0) +  (q^2 / \mu^2)^2 \Pi^{(2)}_0 (0) + O_0(q^6) \Bigr]
\nonumber\\
&=& - {\Delta^2_1 \Delta^2_2 \over q^4} \Pi_0(0) - {\Delta^2_1
\over q^2} a_2 \Pi^{(1)}_0 (0) - a_1a_2\Pi^{(2)}_0 (0)- O(q^2).
\end{eqnarray}
Again the first and second terms should be shifted to the third
type of terms and combined with terms containing there the same
powers of $1/q^2$, while the last two terms should be shifted to
the first type of terms.

Similarly to the Taylor expansion (4.12), one has

\begin{equation}
\Pi_m(q^2)\Pi_n (q^2)= \Pi_{mn} (q^2) = \Pi_{mn}(0) + (q^2 /
\mu^2) \Pi^{(1)}_{mn} (0) + (q^2 / \mu^2)^2 \Pi^{(2)}_{mn} (0) +
O_{mn}(q^6).
\end{equation}
Then, for example the mixed term $(-1/q^2) \Pi_0(q^2)
\Pi_1(q^2)\Delta^2_2$ can be split as

\begin{eqnarray}
-{ \Delta^2_2 \over q^2} \Pi_0(q^2) \Pi_1(q^2) &=& -{ \Delta^2_2
\over q^2} \Bigl[ \Pi_{01}(0) + (q^2 / \mu^2) \Pi^{(1)}_{01} (0) +
O_{01}(q^4) \Bigr] \nonumber\\
&=& -{ \Delta^2_2 \over q^2} \Pi_{01}(0) - a_2 \Pi^{(1)}_{01} (0)
+ O(q^2),
\end{eqnarray}
so again the first term should be shifted to the third type of
terms and combined with the terms containing the corresponding
powers of $1/q^2$, while other terms are to be shifted to the
first type of terms.

Completing this exact splitting/shifting procedure in the
expansion (4.11), one can in general represent it as follows:

\begin{equation}
d(q^2) = \Bigl( {\Delta^2 \over q^2} \Bigr) B_1(\lambda, \alpha,
\xi, g^2) + \Bigl( {\Delta^2 \over q^2} \Bigr)^2 B_2(\lambda,
\alpha, \xi, g^2) + \Bigl( {\Delta^2 \over q^2} \Bigr)^3
B_3(\lambda, \alpha, \xi, g^2) + f_3(q^2) + ....,
\end{equation}
where we used notations (4.2), since the coefficients of the
above-used Taylor expansions depend in general on the same set of
parameters: $\lambda, \alpha, \xi, g^2$. The invariant function
$f_3(q^2)$ is dimensionless and regular at small $q^2$; otherwise
it remains arbitrary. The generalization on the next iterations is
almost obvious. Let us only note that in this case more terms in
the corresponding Taylor expansions should be kept "alive".

\subsection{The exact structure of the general iteration solution}

Substituting the generalization of the expansion (4.17) on all
iterations and omitting the tedious algebra, the general iteration
solution for the regularized full gluon propagator (3.10) can be
exactly decomposed as the sum of the two principally different
terms as follows:

\begin{eqnarray}
D_{\mu\nu}(q; \Delta^2) = D^{INP}_{\mu\nu}(q; \Delta^2)+
D^{PT}_{\mu\nu}(q) &=& i T_{\mu\nu}(q) {\Delta^2 \over (q^2)^2}
\sum_{k=0}^{\infty} (\Delta^2 / q^2)^k \sum_{m=0}^{\infty}
\Phi_{k,m}(\lambda, \alpha,
\xi, g^2) \nonumber\\
&+& i \Bigr[ T_{\mu\nu}(q) \sum_{m=0}^{\infty} A_m(q^2) + \xi
L_{\mu\nu}(q) \Bigl] {1 \over q^2},
\end{eqnarray}
where the superscript "INP" stands for the intrinsically NP part
of the full gluon propagator. We distinguish between the two terms
in Eq. (4.18) by the explicit presence of the mass gap and the
character of the corresponding IR singularities (see below). Let
us emphasize that the general problem of convergence of the
formally regularized series (4.18) is irrelevant here. Anyway, the
problem how to remove all types of the UV divergences (overlapping
\cite{11} and overall \cite{1,2,6,7,8,9}) is a standard one. Our
problem will be how to deal with severe IR singularities due to
their novelty and genuine NP character. Fortunately, there already
exists a well-elaborated mathematical formalism for this purpose,
namely the distribution theory (DT) \cite{12} into which the
dimensional regularization method (DRM) \cite{13} should be
correctly implemented (see also Refs. \cite{10,14}).

As mentioned above, we distinguish between the two terms in the
full gluon propagator (4.18) by the explicit presence of the mass
gap (when it formally goes to zero then the PT term survives
only). This is the first necessary condition. The second equally
necessary condition is the nature of the corresponding IR
singularities. The INP part of the full gluon propagator is
characterized by the presence of severe power-type (or
equivalently NP) IR singularities $(q^2)^{-2-k}, \ k=0,1,2,3,...$.
So these IR singularities are defined as more singular than the
power-type IR singularity of the free gluon propagator
$(q^2)^{-1}$, which thus can be defined as the PT IR singularity.
The INP part depends only on the transversal degrees of freedom of
gauge bosons. Though its coefficients $\Phi_{k,m}(\lambda, \alpha,
\xi, g^2)$ may explicitly depend on the gauge-fixing parameter
$\xi$, the structure of this expansion itself does not depend on
it. The INP part of the full gluon propagator in Eq. (4.18) is
nothing but the corresponding Laurent expansion in integer powers
of $q^2$ accompanied by the corresponding powers of the mass gap
squared and multiplied by the sum over the $q^2$-independent
factors, the so-called residues $\Phi_k(\lambda, \alpha, \xi, g^2)
= \sum_{m=0}^{\infty} \Phi_{k,m}(\lambda, \alpha, \xi, g^2)$. The
sum over $m$ indicates that an infinite number of iterations (all
iterations) of the corresponding regularized skeleton loop
integrals invokes each severe IR singularity labelled by $k$. It
is worth emphasizing that now this Laurent expansion cannot be
summed up into anything similar to the initial Eq. (3.9), since
its residues at poles gain additional contributions due to the
splitting/shifting procedure, i.e., they become arbitrary.
However, this arbitrariness is not a problem, because severe IR
singularities should be treated by the DRM correctly implemented
into the DT. For this the dependence of the residues on their
arguments is all that matters and not their concrete values.

The PT part of the full gluon propagator, which has only the PT IR
singularity, remains undetermined. In the PT part the sum over $m$
again indicates that all iterations contribute to the PT gluon
form factor $d^{PT}(q^2) = \sum_{m=0}^{\infty} A_m(q^2)$. What we
know about $A_m(q^2)$ functions is only that they are regular
functions at small $q^2$; otherwise they remain arbitrary, but
$d^{PT}(q^2)$ should satisfy AF at large $q^2$. This is the price
we have paid to fix exactly the functional dependence of the INP
part of the full gluon propagator. Anyway, just this part gives
rise to the dominant contributions to the numerical values of
physical quantities in low-energy QCD (see discussion below as
well).

Both terms in Eq. (4.18) are valid in the whole energy/momentum
range, i.e., they are not asymptotics. At the same time, we have
achieved the exact separation between the two terms responsible
for the NP (dominating in the IR ($q^2 \rightarrow 0$)) and the
nontrivial PT (dominating in the UV ($q^2 \rightarrow \infty$))
dynamics in the true QCD vacuum. It is worth emphasizing once more
that we exactly distinguish between the two terms in Eq. (4.18) by
the character of the corresponding IR singularities. This
necessary condition includes the existence of a special
regularization expansion for severe (i.e., NP) IR singularities,
while for the PT IR singularity it does not exist \cite{10,12}
(see Appendix B as well). Due to the character of the IR
singularity the longitudinal component of the full gluon
propagator should be included into its PT part, so its INP part
becomes automatically transversal. In Refs. \cite{10,14,15} we
came to the same structure (4.18) but in a rather different way.

Thus the true QCD vacuum is really beset with severe IR
singularities. Within the general iteration solution they should
be summarized (accumulated) into the full gluon propagator and
effectively correctly described by its structure in the deep IR
domain, exactly represented by its INP part. The second step is to
assign a mathematical meaning to the integrals, where such kind of
severe IR singularities will explicitly appear, i.e., to define
them correctly in the IR region \cite{10,15}. Just this violent IR
behavior makes QCD as a whole an IR unstable theory, and therefore
it may have no IR stable fixed point, indeed \cite{1}, which means
that QCD itself might be a confining theory without involving some
extra degrees of freedom \cite{16,17,18,19,20,21,22,23}. In
section VI below we will show that this is so, indeed.

In summary, the above-mentioned separation between the INP and PT
terms in the full gluon propagator (4.18) is not only exact but
unique as well. The general iteration solution (4.18) is
inevitably severely singular in the IR limit ($q^2 \rightarrow
0$), and this does not depend on the special gauge choice.

\subsection{Remarks on overlapping divergences}

The mass gap which appears first in the gluon SD equation (3.6) is
the main object we were worried about to demonstrate explicitly
its crucial role within our approach. Let us make, however, a few
remarks in advance. As it follows from the standard gluon SD
equation (3.6), the corresponding equation for the gluon
self-energy looks like

\begin{equation}
D^{-1}(q) = D^{-1}_0(q) - q^2 \Pi(q^2; D) - \Delta^2(\lambda; D),
\end{equation}
where we omit, for simplicity, the tensor indices, as well as the
longitudinal part of the subtracted gluon self-energy. In order to
unravel overlapping UV divergence problems in QCD, the necessary
number of the differentiation with respect to the external
momentum should be done first (in order to lower divergences).
Then the point-like vertices, which are present in the
corresponding skeleton loop integrals should be replaced by their
full counterparts via the corresponding integral equations.
Finally, one obtains the corresponding SD equations which are much
more complicated than the standards ones, containing different
scattering amplitudes. These skeleton expansions are, however,
free from the above-mentioned overlapping divergences. Of course,
the real procedure (\cite{11} and references therein) is much more
tedious than briefly described above. However, even at this level
it is clear that by taking derivatives with respect to the
external momentum $q$ in the SD equation for the gluon self-energy
(4.19), the main initial information on the mass gap will be
totally lost. Whether it will be somehow restored or not at the
later stages of the renormalization program is not clear at all.
Thus in order to remove overlapping UV divergences ("the water")
from the SD equations and skeleton expansions, we are in danger to
completely lose the information on the dynamical source of the
mass gap ("the baby") within our approach. In order to avoid this
danger and to be guaranteed that no dynamical information are
lost, we are using the standard gluon SD equation (3.6). The
presence of any kind of UV divergences (overlapping and usual
(overall)) in the skeleton expansions will not cause any problems
in order to detect the mass gap responsible for the IR structure
of the true QCD vacuum. In other words, the direct iteration
solution of the standard gluon SD equation (3.6) or equivalently
(3.7) is reliable to realize a mass gap, and thus to make its
existence perfectly clear. The problem of convergence of such
regularized skeleton loop series which appear in Eq. (4.18) is
completely irrelevant in the context of the present investigation.
Anyway, we keep any kind of UV divergences under control within
our method, since we are working with the regularized quantities.
At the same time, the existence of a mass gap responsible for the
IR structure of the full gluon propagator does not depend on
whether overlapping divergences are present or not in the SD
equations and corresponding skeleton expansions. As argued above,
the existence of a mass gap is only due to the TVCGI/S. All this
is the main reason why our starting point is the standard gluon SD
equation (3.6) for the unrenormalized (but necessarily
regularized) Green's functions (this also simplifies notations).
For some preliminary discussion of the renormalization of the
regularized mass gap see section VI below.

\section{Restoration of the transversality of gauge bosons}

Many important quantities in QCD, such as the gluon and quark
condensates, the topological susceptibility, the Bag constant (see
discussion below), etc., are defined only beyond the PT
\cite{24,25,26}. This means that they are determined by such
$S$-matrix elements (correlation functions) from which all types
of the PT contributions should be, by definition, subtracted. It
is worth emphasizing that these subtractions are inevitable also
for the sake of self-consistency. In low-energy QCD there exist
relations between different correlation functions, for example,
the Witten-Veneziano (WV) and Gell-Mann-Oakes-Renner (GMOR)
formulae. The former \cite{27,28} relates the pion decay constant
and the mass of the $\eta'$ meson to the topological
susceptibility. The latter \cite{25,29} relates the chiral quark
condensate to the pion decay constant and its mass. The famous
trace anomaly relation (see, for example Refs. \cite{25,28} and
references therein) relates the above-mentioned Bag constant to
the gluon and quark condensates. Defining thus the topological
susceptibility and the gluon and quark condensates by the
subtraction of all types of the PT contributions, it would not be
self-consistent to retain them in the correlation function,
determining the pion decay constant, and in the expressions for
the pion and $\eta'$ meson masses.

Anyway, to calculate correctly any truly NP quantity from first
principles in low-energy QCD one has to begin with making
subtractions at the fundamental quark-gluon level. So let us
briefly formulate here a general method how to make the gluon
propagator relevant for the NP QCD to be automatically free of the
PT contributions (for more detailed formulation see Ref.
\cite{30}). Using the exact decomposition (4.18), let us define
the INP gluon propagator by the corresponding subtraction as
follows:

\begin{equation}
D^{INP}_{\mu\nu}(q; \Delta^2) = D_{\mu\nu}(q; \Delta^2)-
D_{\mu\nu}(q; \Delta^2=0) = D_{\mu\nu}(q; \Delta^2)-
D^{PT}_{\mu\nu}(q),
\end{equation}
so that the full gluon propagator becomes an exact sum of the two
different terms

\begin{equation}
D_{\mu\nu}(q; \Delta^2) = D^{INP}_{\mu\nu}(q; \Delta^2) +
D^{PT}_{\mu\nu}(q),
\end{equation}
in complete agreement with Eq. (4.18). The principal difference
between the full gluon propagator $D_{\mu\nu}(q; \Delta^2)$ and
the INP gluon propagator $D^{INP}_{\mu\nu}(q; \Delta^2)$ is that
the latter one is free of the PT contributions, while the former
one, being also NP, is "contaminated" by them. Also, the INP gluon
propagator is manifestly transversal, i.e., does not depend
explicitly on the gauge-fixing parameter (see below). This results
in the fact that the longitudinal term of the full gluon
propagator is contained in its PT part, because of the restoration
of the color gauge invariance/symmetry at the final stage within
our approach. So after subtraction it is cancelled in the INP
gluon propagator, which otherwise cannot not be really
transversal. Since the formal PT limit $\Delta^2=0$ is uniquely
defined in our method, the separation between the INP and PT gluon
propagators is uniquely defined as well (let us note that the mass
gap is either finite or zero--the PT limit--, i.e., it cannot be
infinitely large at fixed $\lambda$). In general we distinguish
between the two different phases (the INP and PT ones) in QCD not
by the strength of the coupling constant, but by the presence of a
mass gap (in this case the coupling constant plays no any role as
it follows from our consideration).

Evidently, the subtraction (5.1) is equivalent to the subtraction
made at the level of the full gluon form factor in Eq. (3.10) as
follows:

\begin{equation}
d(q^2) = d(q^2) - d^{PT}(q^2)+ d^{PT}(q^2)= d^{INP} (q^2) +
d^{PT}(q^2).
\end{equation}
It is worth emphasizing once more, that making the above-defined
subtraction, we are achieving the two goals simultaneously: the
transversality of the gluon propagator relevant for the truly NP
QCD, and it automatically becomes free of the PT contributions
("PT contaminations") as well. So our prescription for the
subtraction at the fundamental gluon level is simply reduced to
the replacement of the general iteration solution by its INP part
everywhere, i.e.,

\begin{equation}
D_{\mu\nu}(q; \Delta^2) \longrightarrow D^{INP}_{\mu\nu}(q;
\Delta^2),
\end{equation}
and/or equivalently

\begin{equation}
d(q^2; \Delta^2) \rightarrow d^{INP} (q^2; \Delta^2).
\end{equation}
Their explicit expressions are given below. $D^{INP}_{\mu\nu}(q;
\Delta^2)$ is the full gluon propagator, but free of the PT
contributions. In addition, it is manifestly transversal. $d^{INP}
(q^2; \Delta^2)$ is also free of the PT contributions. It can be
considered as the INP effective charge, i.e., $d^{INP} (q^2;
\Delta^2) \equiv \alpha^{INP}_s (q^2; \Delta^2)$, not losing
generality. In other words, the replacements (5.4) and/or (5.5)
are necessary to be made first at the fundamental gluon level in
order to correctly calculate from first principles any truly NP
physical quantities and processes in low-energy QCD as emphasized
above.

The necessity of such kind of the subtraction and other types ones
has been discussed and justified in our paper \cite{30} (see also
references therein), where some concrete examples are present as
well. Let us discuss here in more detail one rich example of these
subtractions. One of the main characteristics of the true QCD
ground state is the Bag constant. It is just defined as the
difference between the PT and NP vacuum energy densities (VEDs).
So, we can symbolically put $B = VED^{PT} - VED$, where $VED$ is
the NP but "contaminated" by the PT contributions (i.e., this is a
full $VED$ like the full gluon propagator). At the same time, in
accordance with our method we can continue as follows: $B =
VED^{PT} - VED = VED^{PT} - [VED - VED^{PT} + VED^{PT}] = VED^{PT}
- [VED^{NP} + VED^{PT}] = - VED^{NP} > 0$, since the VED is always
negative. Thus the Bag constant is nothing but the truly NP VED,
apart from the sign, by definition, and thus is completely free of
the PT "contaminations". Symbolic subtraction presented here
includes the subtraction at the fundamental gluon level, i.e.,
relation (5.3), and two others at the hadronic level, i.e., when
the gluon degrees of freedom should be integrated out. For a
concrete detailed procedure of how to define correctly and
actually calculate the Bag constant from first principles by
making all necessary subtractions at all levels on the basis of
our solution for the INP QCD see Ref. \cite{31}, where the
relation between the Bag constant and gluon condensate is
explicitly shown as well.

Concluding, the first necessary subtraction (5.1) makes the full
gluon propagator, relevant for the NP QCD, automatically free of
the PT contributions ("PT contaminations"), as well as purely
transversal. The general role of ghost degrees of freedom is to
cancel the longitudinal term in the full gluon propagator in every
order of the PT, going thus beyond it. Though not compromising
their general role in the PT QCD, from our consideration,
nevertheless, it clearly follows that we do not need them for this
purpose. However, this does not mean that we need no ghosts at
all. Of course, we need them in other sectors of QCD, for example
in the quark-gluon ST identity, which contains the so-called
ghost-quark scattering kernel explicitly \cite{1}. It possesses an
important piece of information on quark degrees of freedom
themselves. It should be taken into account in order to correctly
derive confining quark propagator (for preliminary derivation see
Ref. \cite{32} and references therein).

\section{The confining potential and renormalization of the mass gap}

Thus the full gluon propagator which is relevant for the NP QCD
within our approach is as follows:

\begin{equation}
D_{\mu\nu}(q; \Delta^2) = i T_{\mu\nu}(q) d(q^2; \Delta^2) {1
\over q^2} = i T_{\mu\nu}(q) {\Delta^2 \over (q^2)^2} f(q^2),
\end{equation}
where

\begin{eqnarray}
f(q^2) &=& \sum_{k=0}^{\infty} (\Delta^2 / q^2)^k \Phi_k(\lambda,
\alpha, \xi, g^2), \nonumber\\
\Phi_k(\lambda, \alpha, \xi, g^2) &=& \sum_{m=0}^{\infty}
\Phi_{k,m}(\lambda, \alpha, \xi, g^2),
\end{eqnarray}
and the effective charge in this case is to be defined as follows:

\begin{equation}
d(q^2; \Delta^2) \equiv \alpha_s(q^2; \Delta^2) = {\Delta^2 \over
q^2} f(q^2).
\end{equation}
Evidently, after making the above-described subtractions or
equivalently the replacements (5.4) and (5.5), we can omit the
superscript "INP" (this simplifies notations).

Let us recall some remarkable features of the full gluon
propagator (6.1) (for additional discussion see the
above-mentioned our paper \cite{30}). First of all, it depends
only on the transversal degrees of freedom of gauge bosons, while
the non-explicit dependence on the gauge-fixing parameter $\xi$
remains, of course. Also, its functional dependence is uniquely
fixed up to the expressions for the residues $\Phi_k(\lambda,
\alpha, \xi, g^2)$, and it is valid in the whole energy/momentum
range. At large momentum it looks formally as an Operator Product
Expansion (OPE) of the gluon propagator. However, it is suppressed
in this limit ($q^2 \rightarrow \infty$), and thus it is free of
the above-discussed overlapping divergences, which plague the
dominating in this limit the PT part of the full gluon propagator.
It explicitly depends on the mass gap, so that when it formally
goes to zero this solution vanishes, as well as it is free of the
PT "contaminations" due to the character of severe IR
singularities. No approximation/truncations and no special gauge
choice have been made for the corresponding regularized constant
skeleton loop integrals, contributing to the residues
$\Phi_k(\lambda, \alpha, \xi, g^2)$ over all iterations, labelled
by $m$ in Eq. (6.2). The full gluon propagator (6.1) has a
drastically different behavior in the deep IR region ($q^2
\rightarrow 0$) from the free gluon propagator, which is
determined by the explicit presence of the mass gap.

\subsection{The Weierstrass-Sokhocky-Kazorati theorem}

A new surprising feature of this solution is that its both
asymptotics at zero ($q^2 \rightarrow 0$) and at infinity ($q^2
\rightarrow \infty$) are to be determined by its $(q^2)^{-2}$
structure only. This structure determines the behavior of the
solution (6.1) at infinity, since all other terms in the expansion
(6.1) are suppressed in this limit, and the behavior of this
structure at infinity is not dangerous. So the main problem with
our solution (6.1) is its structure in the deep IR region ($q^2
\rightarrow 0$). Fortunately, there exist two mathematical
theories which are of a great help in this case. We have already
mentioned the DT which should be complemented by the DRM. However,
let us begin with the theory of functions of complex variable
\cite{33}, which explains the behavior of our solution (6.1) in
the deep IR region.

The function $f(q^2)$ is defined by its Laurent expansion, and
thus it has an isolated essentially singular point at $q^2=0$. Its
behavior in the neighborhood of this point is regulated by the
Weierstrass-Sokhocky-Kazorati (WSK) theorem which tells that

\begin{equation}
\lim_{n \rightarrow \infty}f(q^2_n) = Z, \quad q^2_n \rightarrow
0,
\end{equation}
where $Z$ is any complex number, and $\{q^2_n \}$ is a sequence of
points $q^2_1, q^2_2, ..., q^2_n$, for which the above-displayed
limit always exists. Of course, $Z$ remains arbitrary (it depends
on the chosen sequence of points), but in general it depends on
the same set of parameters as the residues, i.e., $Z \equiv
Z(\lambda, \alpha, \xi, g^2)$. This theorem thus allows one to
replace the Laurent expansion $f(q^2)$ by $Z$ when $q^2
\rightarrow 0$, i.e.,

\begin{equation}
f(0; \lambda, \alpha, \xi, g^2) \rightarrow Z(\lambda, \alpha,
\xi, g^2).
\end{equation}
There is no doubt that the only real severe (i.e., NP) IR
singularity of the full gluon propagator (6.1) is the $(q^2)^{-2}$
NP IR singularity, while the Laurent expansion $f(q^2)$ should be
treated in accordance with the WSK theorem.

Our consideration at this stage is necessarily formal, since the
mass gap remains unrenormalized yet as well as all other
quantities. So far it has been only regularized, i.e., $\Delta^2
\equiv \Delta^2(\lambda, \alpha, \xi, g^2)$. However, a
preliminary explanation of the renormalization of the mass gap can
be given even at this stage. Due to the above-formulated WSK
theorem, the full gluon propagator (6.1) becomes

\begin{equation}
D_{\mu\nu}(q; \Delta^2) = i T_{\mu\nu}(q) { 1 \over (q^2)^2}
Z(\lambda, \alpha, \xi, g^2) \Delta^2 (\lambda, \alpha, \xi, g^2),
\end{equation}
not losing generality, so just the $(q^2)^{-2}$-structure of the
full gluon propagator (6.1) is all that matters, indeed. Let us
now define the renormalized (R) mass gap as follows:

\begin{equation}
\Delta^2_R =  Z(\lambda, \alpha, \xi, g^2) \Delta^2 (\lambda,
\alpha, \xi, g^2),
\end{equation}
so that we consider $Z(\lambda, \alpha, \xi, g^2)$ as the
multiplicative renormalization constant for the mass gap, and
$\Delta^2_R$ is the physical mass gap within our approach.
Precisely this quantity should be positive (due to the WSK
theorem, we can always choose such $Z$ in order to make
$\Delta^2_R$ positive), finite, gauge-independent, etc. In other
words, it should exist when $\lambda \rightarrow \infty$ and
$\alpha =0$ (for a more detailed consideration see below).

Thus the full gluon propagator relevant for the truly NP QCD (or
equivalently INP QCD) finally becomes

\begin{equation}
D_{\mu\nu}(q; \Delta^2_R) = i T_{\mu\nu}(q) { \Delta^2_R \over
(q^2)^2}.
\end{equation}
The renormalization of the mass gap is an example of the NP
renormalization. The corresponding renormalization constant
$Z(\lambda, \alpha, \xi, g^2)$ appears naturally, so the general
renormalizability of QCD is not affected. Since we were able to
accumulate all the quadratic divergences into the renormalization
of the mass gap, the $(q^2)^{-2}$-type behavior of the relevant
gluon propagator (6.8) at infinity is not dangerous (as mentioned
above), i.e., it cannot undermine the renormalizability of QCD.
Moreover, in the subsequent paper we will show that the INP QCD is
UV finite theory.

As emphasized above, the real problem with our solution (6.8) is
the behavior at the origin ($q^2 \rightarrow 0$), since its IR
singularity represents the so-called severe IR singularity, and
the PT fails to deal with it. Before explaining in general terms
how to correctly treat it within the DT, complemented by the DRM,
let us make a few remarks in advance. It is well known that the
gluon propagator (6.8) satisfies the Wilson criterion of quark
confinement - Area law \cite{34,35}. Equivalently, it leads to the
linear rising potential between heavy (static) quarks \cite{36}
"seen" by lattice QCD as well \cite{37}. That is why we call it
the confining potential. Let us emphasize, however, that our
result (6.8) is exact, so it is neither IR nor UV asymptotic. One
would think that it has a rather simple functional structure but
this is not the case. A special regularization expansion is to be
used in order to deal with its severe IR singularity (see Appendix
B and discussion below).

Before going to the description of the IR multiplicative
renormalization (IRMR) program, it is instructive to find
explicitly the corresponding $\beta$-function. From Eq. (6.8) it
follows that the effective charge is

\begin{equation}
\alpha_s(q^2; \Delta^2_R) = {\Delta^2_R \over q^2}.
\end{equation}
Then from the renormalization group equation for the renormalized
effective charge (6.9), which determines the $\beta$-function,

\begin{equation}
q^2 {d \alpha_s(q^2; \Delta^2_R) \over dq^2} = \beta(\alpha_s(q^2;
\Delta^2_R)),
\end{equation}
it simply follows that

\begin{equation}
\beta(\alpha_s(q^2; \Delta^2_R))= - \alpha_s(q^2; \Delta^2_R).
\end{equation}
Thus, one can conclude that the corresponding $\beta$-function as
a function of its argument is always in the domain of attraction
(i.e., negative). So it has no IR stable fixed point indeed as it
is required for the confining theory \cite{1}.

\subsection{IR multiplicative renormalization (IRMR)}

Let us show now that expression (6.8) is an exact result, and it
is neither IR nor UV asymptotic as underlined above. For this
purpose, it is instructive to begin with the initial expressions
(6.1) and (6.2). Because of the summation over $k$, nothing should
depend on it. This is in agreement with what we already know from
the WSK theorem. Thus the only NP IR singularity of Eq. (6.8) is
its $(q^2)^{-2}$-structure. If $q$ is an independent skeleton loop
variable, then the dimensional regularization of this NP IR
singularity is given by the expansion (B7) at $k=0$, namely

\begin{equation}
(q^2)^{- 2} = { 1 \over \epsilon} \Bigr[ a(0)\delta^4(q) +
O(\epsilon) \Bigl], \quad \epsilon \rightarrow 0^+,
\end{equation}
and $a(0) = \pi^2$ (see Appendix B). Due to the $\delta^4(q)$
function in the residue of this expansion, all the test functions
which appear under corresponding skeleton loop integrals should be
finally replaced by their expression at $q=0$. So Eq. (6.1)
effectively becomes

\begin{equation}
D_{\mu\nu}(q; \Delta^2) = i T_{\mu\nu}(q) {\Delta^2 \over (q^2)^2}
f(0),
\end{equation}
and taking into account the replacement (6.5) (i.e., result of the
WSK theorem) and the definition (6.7), one finally arrives at Eq.
(6.8), indeed. The only problem remaining to solve is how to
remove the pole $1/ \epsilon$ which necessarily appears in the
full gluon propagator. After substituting of the dimensionally
regularized expansion (6.12) into the Eq. (6.8) or equivalently
into the Eq. (6.13), it becomes

\begin{equation}
D_{\mu\nu}(q; \Delta^2_R) = { 1 \over \epsilon} i T_{\mu\nu}(q)
\Delta^2_R \delta^4(q),
\end{equation}
where we include $\pi^2$ into the renormalized mass gap and
retain, for convenience, the same notation. For simplicity, the
terms of the order $O(\epsilon)$ are not shown.

As emphasized in Appendix B, in the presence of the IR
regularization parameter $\epsilon$ all the Green's functions and
parameters depend in general on it. The only way to remove the
pole in $\epsilon$ from the full gluon propagator (6.14) is to
define the IR renormalized mass gap as follows:

\begin{equation}
\Delta^2_R = X(\epsilon) \bar{\Delta}^2_R = \epsilon
\bar{\Delta}^2_R, \quad \epsilon \rightarrow 0^+,
\end{equation}
where $X(\epsilon) = \epsilon$ is the IRMR constant for the mass
gap, and the IR renormalized mass gap $\bar{\Delta}^2_R$ exists as
$\epsilon \rightarrow 0^+$, by definition, contrary to
$\Delta^2_R$. In both expressions for the mass gap the dependence
on $\epsilon$ is assumed but not shown explicitly. Evidently,
after the IR renormalization of the mass gap, the terms of the
order $O(\epsilon)$ become the terms of the order $O(\epsilon^2)$.
After going to the IR renormalized quantities they can be omitted
from the consideration without any problems.

The important observation is that the renormalization of the mass
gap automatically IR renormalizes the full gluon propagator as
well, i.e., its IRMR constant is $X(\epsilon) = \epsilon$, so
$D_{\mu\nu}(q) = X(\epsilon)\bar{D}_{\mu\nu}(q)$. The IR and UV
renormalized gluon propagator becomes

\begin{equation}
D_{\mu\nu}(q; \bar{\Delta}^2_R) = i T_{\mu\nu}(q) \bar{\Delta}^2_R
\delta^4(q),
\end{equation}
where we introduce the notation $\bar{D}_{\mu\nu}(q) \equiv
D_{\mu\nu}(q; \bar{\Delta}^2_R)$. In a subsequent paper we will
show that the IR renormalizaion of the full gluon propagator or
equivalently of the mass gap is completely sufficient to remove
all severe IR singularities (parameterized in terms of the IR
regularization parameter $\epsilon$ and introduced within the DT,
complemented by DRM) from all the multi-loop skeleton integrals
which may appear in the INP QCD. So this theory is IR
renormalizable as well. However, let us note in advance that
beyond one-loop skeleton integrals the analysis should be done in
a more sophisticated way. Instead of the $\delta$ function in the
residues its derivatives will appear. They should be treated in
the sense of the DT. Fortunately, as mentioned in Appendix B, the
IR renormalization of the theory is not undermined, since a pole
in $\epsilon$ is always a simple pole $1 / \epsilon$ for each
independent skeleton loop variable (see expansion (B7)).

\subsection{The general criterion of gluon confinement}

Let us now substitute the definition (6.15) into the gluon
propagator (6.8), i.e, express it in terms of the IR renormalized
mass gap. Then one obtains

\begin{equation}
D_{\mu\nu}(q; \bar{\Delta}^2_R) = \epsilon \times i T_{\mu\nu}(q)
{\bar{\Delta}^2_R \over (q^2)^2}, \quad \epsilon \rightarrow 0^+.
\end{equation}
Due to the distribution nature of severe IR singularities, the two
principally different cases should be separately considered.

1). If the gluon momentum $q$ is an independent skeleton loop
variable, then, as emphasized repeatedly above, the initial
$(q^2)^{-2}$ severe IR singularity should be regularized with the
help of the expansion (6.12). Finally one arrives at Eq. (6.16) in
the $\epsilon \rightarrow 0^+$ limit as it should be. It makes
sense to note here that if under the integrals the effective
charge (6.9) is multiplied by at least $(q^2)^{-1}$, then the IR
renormalization of the mass gap is needed, otherwise -- not.

2). If, however, the gluon momentum $q$ is not a skeleton loop
variable (i.e., it is external momentum), then the initial
$(q^2)^{-2}$ severe IR singularity cannot be treated as the
distribution, i.e., the regularization expansion (6.12) is not the
case to use. The function $(q^2)^{-2}$ is a standard one, and the
gluon propagator (6.17) vanishes as $\epsilon$ goes to zero, i.e.,

\begin{equation}
D_{\mu\nu}(q; \bar{\Delta}^2_R) = \epsilon \times i T_{\mu\nu}(q)
{\bar{\Delta}^2_R \over (q^2)^2} \sim \epsilon, \quad \epsilon
\rightarrow 0^+.
\end{equation}
It is worth emphasizing that the final $\epsilon \rightarrow 0^+$
limit is permitted to take only after expressing all the Green's
functions and parameters in terms of their IR renormalized
counterparts because they, by definition, exist in this limit.
This behavior is gauge-invariant and prevents the transversal
gluons to appear in asymptotic states. We consider it as the exact
criterion of gluon confinement, and thus color gluons can never be
isolated.  For the first time it has been obtained in Ref.
\cite{10}.

Concluding, in multi-loop skeleton diagrams containing gluon
propagators or external gluon legs the starting expression for
them within our approach is always Eq. (6.17). If $q$ is an
independent skeleton loop variable, then one arrives at Eq.
(6.16), otherwise at Eq. (6.18). As mentioned above, the
derivatives of the $\delta$ function may also appear when the
number of independent skeleton loop variables does not coincide
with the number of the gluon propagators (more detailed analysis
has to be done).

\subsection{The general criterion of quark confinement}

It is instructive to formulate here the quark confinement
criterion as well in advance. It consists of two independent
conditions.

1). The $first \ necessary$ condition, formulated at the
fundamental (microscopic) quark-gluon level, is the absence of the
pole-type singularities in the quark Green's function at any gauge
on the real axe at some finite point in the complex momentum
space, i.e.,

\begin{equation}
S(p) \neq { Z_2 \over \hat p - m_{ph}},
\end{equation}
where $Z_2$ is the standard quark wave function renormalization
constant, while $m_{ph}$ is the mass to which a physical meaning
could be assigned. In other words, the quark always remains an
off-mass-shell object. Such an understanding (interpretation) of
quark confinement comes apparently from Gribov's approach to quark
confinement \cite{38} and Preparata's massive quark model (MQM) in
which external quark legs were approximated by entire functions
\cite{39}. A quark propagator may or may not be an entire
function, but in any case the pole of the first order (like the
electron propagator has in QED) should disappear (see, for example
Refs. \cite{32,40} and references therein).

2). The $second \ sufficient$ condition, formulated at the
hadronic (macroscopic) level, is the existence of the discrete
spectrum only (no continuum) in bound-states, in order to prevent
quarks to appear in asymptotic states. This condition comes
apparently from 't Hooft's model for two-dimensional QCD with
$N_c$ large limit \cite{41} (see also Refs. \cite{16,32}).

At nonzero temperature and density, for example in quark-gluon
plasma (QGP), the bound-states will be dissolved, so the second
sufficient condition does not work anymore. However, the first
necessary condition remains always valid, of course. In other
words, by increasing temperature or density there is no way to put
quarks on the mass-shell. So what is known as the Deconfinement
phase transition in QGP is in fact the Dehadronization phase
transition.

This definition of quark confinement in the momentum space is
gauge-invariant, flavor independent, etc., and thus it is a
general one. At the same time, the above-mentioned quark
confinement criterion formulated in the configuration space - Area
law (and equivalently the linear rising potential) - is relevant
only for heavy quarks.

\subsection{Physical limits}

We introduced the renormalized mass gap (6.7), defining its
existence when the dimensionless UV regulating parameter $\lambda$
goes to infinity, i.e., in the $\lambda \rightarrow \infty$ limit.
However, nothing was said about the behavior of the coupling
constant squared $g^2$ in this limit. In general it may also
depend on $\lambda$, becoming thus the so-called "running"
effective charge $g^2 \sim \alpha_s \equiv \alpha_s(\lambda)$.
Evidently, nothing will be changed in our results if we choose the
subtraction point $\alpha$ at zero, i.e., put $\alpha=0$. We also
omit the dependence on the gauge-fixing parameter $\xi$ as
unimportant for future discussion. Thus the definition (6.7) can
be written down in the generalized version as follows:

\begin{equation}
M^2 =  Z(\lambda, \alpha_s(\lambda)) \Delta^2 (\lambda,
\alpha_s(\lambda)),
\end{equation}
where we introduce the auxiliary finite mass squared parameter
$M^2$. Since the regularized effective charge depends on
$\lambda$, all the possible types of its behavior in the $\lambda
\rightarrow \infty$ limit should be considered independently from
each other.

1). If $\alpha_s(\lambda) \rightarrow \infty$ as $\lambda
\rightarrow \infty$, then one recovers the strong coupling regime.
It is nothing but the INP phase within our approach. Evidently,
just this limit has been defined as the renormalized mass gap
(6.7), i.e., in fact

\begin{equation}
M^2 =  Z(\lambda, \alpha_s(\lambda))\Delta^2 (\lambda,
\alpha_s(\lambda)) = \Delta^2_R, \quad \lambda \rightarrow \infty,
\quad \alpha_s(\lambda) \rightarrow \infty.
\end{equation}

2). If $\alpha_s(\lambda) \rightarrow c$ as $\lambda \rightarrow
\infty$, where $c$ is a finite constant, then it can be put to
one, not losing generality. This means that the effective charge
becomes unity, and this is only possible for the free gluon
propagator. However, the free gluon propagator contains none of
mass scale parameters, so $M^2$ in this limit is to be zero, i.e.,
in fact

\begin{equation}
M^2 =  Z(\lambda, \alpha_s(\lambda))\Delta^2 (\lambda,
\alpha_s(\lambda)) =0, \quad \lambda \rightarrow \infty, \quad
\alpha_s(\lambda) \rightarrow 1.
\end{equation}

3). If $\alpha_s(\lambda) \rightarrow 0$ as $\lambda \rightarrow
\infty$, then one recovers the weak coupling regime. It is nothing
but the PT phase within our approach. Evidently, this limit has to
be defined as $\Lambda^2_{QCD}$, i.e., in fact

\begin{equation}
M^2 =  Z(\lambda, \alpha_s(\lambda))\Delta^2 (\lambda,
\alpha_s(\lambda)) = \Lambda^2_{QCD}, \quad \lambda \rightarrow
\infty, \quad \alpha_s(\lambda) \rightarrow 0.
\end{equation}

A few remarks are in order. Due to the above-formulated WSK
theorem, the number $Z$ depends on the chosen sequence of points
along which the gluon momentum squared goes to zero. For the first
and third cases we can choose it in order to provide a finite
different limits, while for the second case it can be chosen in
order to provide a zero limit. Thus these three different regimes
are in agreement with the theory of functions of complex variable
\cite{33}. The PT phase is determined by the subtracted part of
the gluon self-energy, which may be only logarithmic divergent as
emphasized above. So $\Lambda^2_{QCD} \equiv \Lambda_{PT}^2$ may
effectively appear under logarithms only, leading finally to AF.
In Ref. \cite{42} it has been noticed that being numerically a few
hundred $MeV$ only, $\Lambda^2_{QCD}$ cannot survive in the UV
limit, so none of the finite mass scale parameters can be
determined by the PT QCD. It should come from the IR, being thus
NP by origin as was just described above.

Our mass gap $\bar{\Delta}^2_R$ determines the power-type
deviation of the full gluon propagator from the free one in the IR
limit ($q^2 \rightarrow 0$). This region (small $q^2$) is
interesting for all the NP effects in QCD, first of all color
confinement. This once more emphasizes the close intrinsic link
between the behavior of QCD at large distances and its NP
dynamics. At the same time, asymptotic QCD scale parameter
$\Lambda_{QCD}$ determines much more weaker logarithmic deviation
of the full gluon propagator from the free one in the UV limit
($q^2 \rightarrow \infty$). From our consideration it follows that
both effects (AF and color confinement) are due to the existence
of the mass gap in QCD. In QED neither AF nor confinement takes
place because there is no mass gap at all. There is no doubt left
that due to renormalization quadratic divergences parameterized as
the regularized mass gap $\Delta^2$ may be absorbed in a
redefinition of two various masses leading to the above-mentioned
two physical mass gaps $\Lambda^2_{NP}$ and  $\Lambda^2_{PT}$.

\section{ Discussion}

Let us denote the version of our mass gap which appears in the
$S$-matrix elements for the corresponding physical
quantities/processes in low-energy QCD as $\Lambda^2_{NP}$ (in
principle they may be slightly different, indeed). Then a symbolic
relation between it and the initial mass gap $\Delta^2$ and
$\Lambda^2_{PT}$ could be written as follows:

\begin{equation}
\Lambda^2_{NP} \longleftarrow^{\infty \leftarrow \alpha_s}_{\infty
\leftarrow \lambda} \ \Delta^2 \ { }^{\alpha_s \rightarrow
0}_{\lambda \rightarrow \infty} \longrightarrow  \ \Lambda^2_{PT}.
\end{equation}
Here $\alpha_s$ is obviously the fine structure coupling constant
of strong interactions. The right-hand-side limit is well known as
the weak coupling regime, and we know how to take it within the
renormalization group equations approach \cite{1,2,6,7,8,9}. The
left-hand-side limit can be regarded as the strong coupling
regime, and we hope that we have explained here how to begin to
deal with it, proposing and formulating nonlinear iteration
procedure. However, there is no doubt that the final goal of this
limit, namely, the mass gap $\Lambda_{NP}$ exists, and should be
renormalization group invariant in the same way as
$\Lambda_{QCD}$. It is solely responsible for the large-scale
structure of the true QCD ground state, while $\Lambda_{PT}$ is
responsible for the nontrivial PT dynamics there.

Evidently, a relation of the type (7.1) is only possible due to
the explicit presence of a mass gap in the full gluon propagator
and in the gluon SD equation of motion. It leads to the exact
separation between the truly NP and nontrivial PT parts (phases)
at the level of a single gluon propagator. A possible relation
between these two phases shown in Eq. (7.1) is a manifestation
that "the problems encountered in perturbation theory are not mere
mathematical artifacts but rather signify deep properties of the
full theory" \cite{43}. The message that we are trying to convey
is that the nontrivial PT phase in the full gluon propagator
indicates the existence of the truly NP one (the INP one within
the general iteration solution) and the other way around.

A few years ago Jaffe and Witten have formulated the following
theorem \cite{5}:

\vspace{3mm}

 {\bf Yang-Mills Existence And Mass Gap:} Prove that
for any compact simple gauge group $G$, quantum Yang-Mills theory
on $\bf{R}^4$ exists and has a mass gap $\Delta > 0$.

\vspace{3mm}

Of course, to prove the existence of the Yang-Mills (YM) theory
with compact simple gauge group $G$ is a formidable task yet. It
is rather a mathematical than a physical problem. However, our
main results obtained in this work can the be formulated similar
to the above-mentioned Jaffe-Witten (JW) theorem as follows:

\vspace{3mm}

{\bf Mass Gap Existence:} If quantum Yang-Mills theory with
compact simple gauge group $G=SU(3)$ exists on $\bf{R}^4$, then it
has a mass gap and confines gluons.

\vspace{3mm}

From the JW presentation of their theorem it clearly follows that
their mass gap is to be identified with our mass gap, after
subtracting all types of the PT contributions at all levels in the
$S$ matrix elements for the corresponding physical
quantities/processes (see symbolic relation (7.1)). On the other
hand, in the AF regime (when all the NP contributions are
suppressed) the mass gap $\Delta^2$ is to be identified with the
asymptotic scale squared (see again symbolic relation (7.1)). QCD
as a formal theory of quark-gluon interactions has a mass gap
$\Delta^2$ which is only regularized, and therefore there is no
guarantee that it is positive. It cannot be related directly to
any of physical quantities/processes. As actual theory of strong
interactions its two different faces come into the play: INP QCD
for low-energy physics with its mass gap $\Lambda^2_{NP}$ and PT
QCD for high-energy physics with its mass gap $\Lambda^2_{PT}
\equiv \Lambda^2_{QCD}$. Both mass gaps have now physical meanings
(they are finite, positive, gauge-invariant, etc.). How $\Delta^2$
becomes either $\Lambda^2_{NP}$ or $\Lambda^2_{PT}$ is already
explained. Apparently, there is no intersection between them.
Similarly to the symbolic relation (7.1), the symbolic relation
between formal QCD and its two physical phases is

\begin{equation}
INP \ QCD \ \Longleftarrow \ QCD \ \Longrightarrow \ PT \ QCD,
\end{equation}
so formal QCD has no physical mass gap, while its two phases do
have as described above. PT QCD is AF, while INP QCD confines
gluons. In the subsequent paper we will show that this theory will
confine quarks as well as will explain spontaneous breakdown of
chiral symmetry. Evidently, it has the mass gap which is only one
responsible for the scale of all the NP effects and processes in
the theory of strong interactions.

\section{Conclusions}

There is no doubt that our approach to the NP QCD will survive
both multiplicative renormalization (MR) programs. In principle,
the UVMR program is not our problem (it is a standard one
\cite{1,2,6,7,8,9,11}). Within the INP solution to QCD our problem
is the IRMR program in order to render the whole theory finite,
i.e., to make it free from all types of severe IR singularities
parameterized in terms of the IR regularization parameter as it
goes to zero at the final stage. It is not a simple task due to
its novelty and really NP character. It requires much more tedious
technical work how to correctly implement the DRM into the DT, and
it is left to be done elsewhere. Anyway, for the sake of
self-consistency it should be done in the framework of the whole
system of the SD equations, complemented by the corresponding ST
identities. However, the UV and IR renormalization of the mass gap
described here will be valid for the renormalization of the
above-mentioned system of equations as well.

It is important to emphasize that the mass gap $\Delta^2$ has not
been introduced by hand. It is hidden in the skeleton loop
integrals, contributing to the full gluon self-energy, and
dynamically generated mainly due to the NL interaction of massless
gluon modes. No truncations/approximations and no special gauge
choice are made for the above-mentioned regularized skeleton loop
integrals. An appropriate subtraction scheme has been applied to
make the existence of a mass gap perfectly clear. Within the
general iteration solution the mass gap shows up explicitly when
the gluon momentum goes to zero. The Lagrangian of QCD does not
contain a mass gap, while it explicitly appears in the gluon SD
equation of motion. This once more underlines the importance of
the investigation of the SD system of equations and identities
\cite{1,2,8,9} for understanding the true structure of the QCD
ground state. We have established the structure of the regularized
full gluon propagator (see Eqs. (3.10) and (3.11)) and the
corresponding SD equation (3.7) in the presence of a mass gap (see
also Appendix A).

In order to realize a mass gap (more precisely its regularized
version), we propose not to impose the transversality condition on
the gluon self-energy, while preserving the color gauge invariance
condition (3.9) for the full gluon propagator. Such a temporary
violation of color gauge invariance/symmetry (TVCGI/S) is
completely NP effect because in the formal PT limit $\Delta^2=0$
this effect vanishes. Since the gluon is not a physical state due
to color confinement, the TVCGI/S in QCD has no direct physical
consequences. None of physical quantites/processes in low-energy
QCD will be directly affected by this proposal.

In QED a mass gap is always in the "gauge prison". It cannot be
realized even temporarily, since the photon is a physical state.
However, in QCD a door of the "color gauge prison" can be opened
for a moment in order to realize a mass gap, because the gluon is
not a physical state. A key to this "door" is the constant
skeleton tadpole term, which explicitly violates the
transversality of the full gluon self-energy \cite{30}. On the
other hand, this "door" can be opened without a key (as any door)
by not imposing the transversality condition on the full gluon
self-energy. So in QED a mass gap cannot be "liberated" from the
vacuum, while photons and electrons can be liberated from the
vacuum in order to be physical states. In QCD a mass gap can be
"liberated" from the vacuum, while gluons and quarks cannot be
liberated from the vacuum in order to be physical states. In other
words, there is no breakdown of $U(1)$ gauge symmetry in QED
because the photon is a physical state. At the same time, a
temporary breakdown of $SU(3)$ color gauge symmetry in QCD is
possible because the gluon is not a physical state (color
confinement).

For the calculations of physical observables from first principles
in low-energy QCD we need the full gluon propagator, which
transversality has been sacrificed in order to realize a mass gap.
However, we have already pointed out how the transversality of the
gluon propagator relevant for the NP QCD is to be restored at the
final stage. In accordance with our prescription it becomes
automatically transversal, free of the PT contributions
("contaminations"), and it regularly depends on the mass gap, so
that it vanishes when the mass gap goes to zero. The role of the
first necessary subtraction (5.1) at the fundamental gluon
propagator level is to be underlined.

Let us emphasize once more that no truncations/approximations and
no special gauge have been made for the corresponding skeleton
loop integrals within our approach, i.e., it is pure NP, by its
nature. So on the general ground we have established that our
solution (6.1), which is relevant for the NP QCD, is always
severely singular in the IR ($q^2 \rightarrow 0$), i.e., the
gluons always remain massless, and this does not depend on the
gauge choice. Moreover, the corresponding IR and UV MR programs
clearly show that its $(q^2)^{-2}$ structure is only important. It
leads to the formulation for the first time of the exact criterion
of gluon confinement (6.18), so it is confining solution, indeed,
see Eq. (6.8). This behavior of the full gluon propagator in
different approximations and gauges has been earlier obtained and
investigated in many papers (see, for example Ref. \cite{10} and
references therein). We have confirmed and thus revitalized these
investigations, in which this behavior has been obtained as an IR
asymptotic solution to the gluon SD equation. {\bf However, let us
emphasize once more that our result is exact and gauge-invariant,
i.e., it is not IR asymptotic.} In other words, it is analytically
proven that the full gluon propagator relevant for the truly NP
QCD (or equivalently INP QCD) and the corresponding effective
charge as a functions of the gluon momentum squared are exactly
given in Eqs. (6.8) and (6.9), respectively. In the
above-mentioned subsequent paper this result will be used in order
to derive confining quark propagator. It is worth noting that in
our previous work \cite{30}, the massive-type solution, leading to
an effective gluon mass, has been also obtained. It becomes smooth
in the Landau gauge only (in this connection see Ref. \cite{44}
and references therein).

In summary, QCD as theory of quark-gluon interactions has a mass
gap $\Delta^2$, possibly realized in accordance with our proposal.
The dynamically generated mass gap is usually related to breakdown
of some symmetry (for example, the dynamically generated quark
mass is an evidence of chiral symmetry breakdown). Here a mass gap
is an evidence of the TVCGI/S. In the presence of a mass gap the
coupling constant plays no role. This is also a direct evidence of
the "dimensional transmutation", $g^2 \rightarrow
\Delta^2(\lambda, \alpha, \xi, g^2)$ \cite{1,45,46}, which occurs
whenever a massless theory acquires masses dynamically. It is a
general feature of spontaneous symmetry breaking in field
theories. At the hadronic level the QCD mass gap $\Delta^2$
undergoes two phase transitions, becoming either $\Lambda_{NP}^2$
responsible for the large-scale QCD dynamics or $\Lambda_{PT}^2$
responsible for its nontrivial short-scale dynamics.
Correspondingly, INP QCD and PT QCD are "two sides of the same
coin"--QCD as a fundamental theory of quark-gluon interactions.

\begin{acknowledgments}

Support in part by HAS-JINR and Hungarian OTKA-T043455 grants (P.
Levai) is to be acknowledged. The author is grateful to A.
Kacharava, N. Nikolaev, P. Forgacs, L. Palla, J. Nyiri, T. Biro
and especially to C. Hanhart and A. Kvinikhidze for useful
discussions and remarks during his stay at IKP (Juelich).

\end{acknowledgments}

\appendix

\section{General structure of the gluon SD equation}

Our strategy is not to impose the transversality condition on the
gluon self-energy in order to realize a mass gap, and only after
to impose the color gauge invariance condition on the full gluon
propagator. To show that this works, it is instructive to
substitute directly the subtracted gluon self-energy (3.5) into
the gluon SD equation (3.6). Then one obtains

\begin{equation}
D_{\mu\nu}(q) = D^0_{\mu\nu}(q) + D^0_{\mu\rho}(q)i[
T_{\rho\sigma}(q) q^2 \Pi^s_1(q^2; D) + q_{\rho}q_{\sigma}
\Pi^s_2(q^2; D)]D_{\sigma\nu}(q) + D^0_{\mu\sigma}(q)i
\Delta^2(\lambda; D) D_{\sigma\nu}(q).
\end{equation}
Let us now introduce the general tensor decompositions of the full
and auxiliary free gluon propagators

\begin{equation}
D_{\mu\nu}(q)=i[T_{\mu\nu}(q) d(q^2) + L_{\mu\nu}(q)d_1(q^2)]{1
\over q^2}
\end{equation}
and

\begin{equation}
D^0_{\mu\nu}(q)=i[ T_{\mu\nu}(q) + L_{\mu\nu}(q) d_0(q^2)]{ 1
\over q^2},
\end{equation}
respectively. The form factor $d_0(q^2)$ introduced into the
unphysical part of the auxiliary free gluon propagator
$D^0_{\mu\nu}(q)$ is needed in order to explicitly show that the
longitudinal part of the subtracted gluon self-energy
$\Pi^s_2(q^2; D)$ plays no role.

The color gauge invariance condition imposed on the full gluon
propagator

\begin{equation}
q_{\mu}q_{\nu}D_{\mu\nu}(q) = i \xi,
\end{equation}
implies $d_1(q^2) = \xi$, so that the full gluon propagator
becomes

\begin{equation}
D_{\mu\nu}(q) = i \left\{ T_{\mu\nu}(q) d(q^2) + \xi L_{\mu\nu}(q)
\right\} {1 \over q^2}.
\end{equation}
Substituting all these decompositions into the gluon SD equation
(A1), one obtains

\begin{equation}
d(q^2) = {1 \over 1 + \Pi^s_1(q^2; D) + (\Delta^2(\lambda; D) /
q^2)},
\end{equation}
and

\begin{equation}
d_0(q^2) = {\xi \over 1 - \xi [\Pi^s_2(q^2; D) +
(\Delta^2(\lambda; D) / q^2)]}.
\end{equation}
However, the auxiliary free gluon propagator defined in Eqs. (A3)
and (A7) is to be equivalently replaced as follows:

\begin{equation}
D^0_{\mu\nu}(q) \Longrightarrow D^0_{\mu\nu}(q) + i \xi
L_{\mu\nu}(q) d_0(q^2) \Bigl[ \Pi^s_2(q^2; D) + {
\Delta^2(\lambda; D) \over q^2} \Bigr] {1 \over q^2},
\end{equation}
where from now on $D^0_{\mu\nu}(q)$ in the right-hand-side is the
standard free gluon propagator, i.e.,

\begin{equation}
D^0_{\mu\nu}(q) = i \left\{ T_{\mu\nu}(q) + \xi L_{\mu\nu}(q)
\right\} {1 \over q^2}.
\end{equation}

Foe further purposes it is convenient to introduce the following
notations:

\begin{equation}
M(q^2) = \Bigl[ \Pi^s_2(q^2; D) + { \Delta^2(\lambda; D) \over
q^2} \Bigr], \quad d_0(q^2) = { \xi \over 1 - \xi M(q^2)},
\end{equation}
so that Eq. (A8) becomes

\begin{equation}
D^0_{\mu\nu}(q) \Longrightarrow D^0_{\mu\nu}(q) + i \xi
L_{\mu\nu}(q) { \xi M(q^2) \over 1 - \xi M(q^2)} {1 \over q^2}.
\end{equation}
Then the gluon SD equation in the presence of the mass gap (A1)
after substitution (A11) and doing some tedious algebra is also to
be equivalently replaced as follows:

\begin{eqnarray}
D_{\mu\nu}(q) &=& D^0_{\mu\nu}(q) + D^0_{\mu\rho}(q)i
T_{\rho\sigma}(q) q^2 \Pi^s_1(q^2; D) D_{\sigma\nu}(q) +
D^0_{\mu\sigma}(q)i \Delta^2(\lambda; D) D_{\sigma\nu}(q)
\nonumber\\
&+& i \xi L_{\mu\nu}(q) {1 \over q^2} \Bigl[ - \xi \Pi^s_2(q^2;D)
+ { \xi M(q^2) \over 1 - \xi M(q^2)} \Bigl[ 1 - \xi \Pi^s_2(q^2;D)
- \xi { \Delta^2(\lambda; D) \over q^2}
 \Bigr] \Bigr], \nonumber\\
\end{eqnarray}
and taking into account the first of notations (A10), one arrives
at

\begin{eqnarray}
D_{\mu\nu}(q) &=& D^0_{\mu\nu}(q) + D^0_{\mu\rho}(q)i
T_{\rho\sigma}(q) q^2 \Pi^s_1(q^2; D) D_{\sigma\nu}(q) \nonumber\\
&+& D^0_{\mu\sigma}(q)i \Delta^2(\lambda; D) D_{\sigma\nu}(q) + i
\xi^2 L_{\mu\nu}(q) { \Delta^2(\lambda; D) \over q^4}.
\end{eqnarray}
Thus the gluon SD equation (A13) does not depend on $d_0(q^2)$ and
$\Pi^s_2(q^2; D)$, i.e., they played their role and then retired
from the scene. Our derivation explicitly shows that the
longitudinal part of the subtracted gluon self-energy
$\Pi^s_2(q^2; D)$ plays no role and can be put formally to zero
without losing generality, and thus making the subtracted gluon
self-energy purely transversal.

Using now the explicit expression for the free gluon propagator
(A9) this equation can be easily simplified to

\begin{equation}
D_{\mu\nu}(q) = D^0_{\mu\nu}(q) - T_{\mu\sigma}(q)
\Bigl[\Pi^s_1(q^2; D) + { \Delta^2(\lambda; D) \over q^2} \Bigr]
D_{\sigma\nu}(q),
\end{equation}
which, of course, coincides with Eq. (3.7), and its "solution" is
again Eq. (A6). Thus, we have established the general structure of
the full gluon propagator (Eqs. (A5) and (A6)) and the
corresponding gluon SD equation (A14), which is equivalent to the
initial Eq. (A1), in the presence of a mass gap.

\section{ IR dimensional regularization within the distribution theory}

In general all the Green's functions in QCD are generalized
functions, i.e., they are distributions. This is especially true
for the NP IR singularities of the full gluon propagator due to
the self-interaction of massless gluons in the QCD vacuum. They
present a rather broad and important class of functions with
algebraic singularities, i.e., functions with nonsummable
singularities at isolated points \cite{12} (at zero in our case).
Roughly speaking, this means that all relations involving
distributions should be considered under corresponding integrals,
taking into account the smoothness properties of the corresponding
space of test functions (for example, $\varphi(q)$ below. Let us
note in advance that in the subsequent paper we will establish the
space in which our generalized functions are continuous linear
functionals).

Let us consider the positively definite ($P>0$) squared
(quadratic) Euclidean form

\begin{equation}
P(q) = q_0^2 +  q_1^2 + q_2^2 + ... + q_{n-1}^2 = q^2,
\end{equation}
where $n$ is the number of the components. The generalized
function (distribution) $P^{\lambda}(q)$, where $\lambda$ being in
general an arbitrary complex number, is defined as

\begin{equation}
(P^{\lambda}, \varphi) = \int_{P>0}P^{\lambda}(q) \varphi(q) d^nq.
\end{equation}
At $Re \lambda \geq 0$ this integral is convergent and is an
analytic function of $\lambda$. Analytical continuation to the
region $Re \lambda < 0$ shows that it has a simple pole at points
\cite{10,12}

\begin{equation}
\lambda = - {n \over 2} - k, \quad k=0, 1, 2 ,3...
\end{equation}

In order to actually define the system of the SD equations in the
deep IR domain, it is necessary to introduce the IR regularization
parameter $\epsilon$, defined as $D = n + 2 \epsilon, \ \epsilon
\rightarrow 0^+$ within a gauge-invariant DRM \cite{13}. As a
result, all the Green's functions and "bare" parameters should be
regularized with respect to $\epsilon$ (see below) which should be
set to zero at the end of the computations. The structure of the
NP IR singularities is then determined (when $n$ is even number)
as follows \cite{12}:

\begin{equation}
(q^2)^{\lambda} = { C_{-1}^{(k)} \over \lambda +(D/2) + k} +
finite \ terms,
\end{equation}
where the residue is

\begin{equation}
 C_{-1}^{(k)} = { \pi^{n/2} \over 2^{2k} k! \Gamma ((n/2) + k) } \times
L^k \delta^n (q)
\end{equation}
with $L = (\partial^2 / \partial q^2_0) + (\partial^2 /
\partial q^2_1) + ... + (\partial^2 / \partial q^2_{n-1})$.

Thus the regularization of the NP IR singularities (B2), on
account of (B3), is nothing but the whole expansion in the
corresponding powers of $\epsilon$ and not the separate term(s).
Let us underline its most remarkable feature. The order of
singularity does not depend on $\lambda$, $n$ and $k$. In terms of
the IR regularization parameter $\epsilon$ it is always a simple
pole $1/ \epsilon$. This means that all power terms in Eq. (B4)
will have the same singularity, i.e.,

\begin{equation}
(q^2)^{- {n \over 2} - k } = { 1 \over \epsilon} C_{-1}^{(k)} +
finite \ terms, \quad \epsilon \rightarrow 0^+,
\end{equation}
where we can put $D=n$ now (i.e., after introducing this
expansion). By "$finite \ terms$" here and everywhere a number of
necessary subtractions under corresponding integrals is understood
\cite{12}. However, the residue at a pole will be drastically
changed from one power singularity to another. This means
different solutions to the whole system of the SD equations for
different set of numbers $\lambda$ and $k$. Different solutions
mean, in their turn, different vacua. In this picture different
vacua are to be labelled by the two independent numbers: the
exponent $\lambda$ and $k$. At a given number of $D(=n)$ the
exponent $\lambda$ is always negative being integer if $D(=n)$ is
an even number or fractional if $D(=n)$ is an odd number. The
number $k$ is always integer and positive and precisely it
determines the corresponding residue at a simple pole, see Eq.
(B3). It would not be surprising if these numbers were somehow
related to the nontrivial topology of the QCD vacuum in any
dimensions.

It is worth emphasizing that the structure of severe IR
singularities in Euclidean space is much simpler than in Minkowski
space, where kinematical (unphysical) singularities due to the
light cone also exist \cite{1,12,47} (in this connection let us
remind that in Euclidean metrics $q^2 = 0$ implies $q_i=0$ and
vice-versa, while in Minkowski metrics this is not so). In this
case it is rather difficult to untangle them correctly from the
dynamical singularities, the only ones which are important for the
calculation of any physical observable. Also, the consideration is
much more complicated in the configuration space \cite{12}. That
is why we always prefer to work in the momentum space (where
propagators do not depend explicitly on the number of dimensions)
with Euclidean signature. We also prefer to work in the covariant
gauges in order to avoid peculiarities of the non-covariant gauges
\cite{1,48,49}, for example, how to untangle the gauge pole from
the dynamical one.

In principle, none of the regularization schemes (how to introduce
the IR regularization parameter in order to parameterize the NP IR
divergences and thus to put them under control) should be
introduced by hand. First of all, it should be well defined.
Secondly, it should be compatible with the DT \cite{12}. The DRM
\cite{13} is well defined, and here we have shown how it should be
introduced into the DT (complemented by the number of
subtractions, if necessary). Though the so-called $\pm i\epsilon$
regularization is formally equivalent to the regularization used
in our paper (see again Ref. \cite{12}), nevertheless, it is
rather inconvenient for practical use. Especially this is true for
the gauge-field propagators, which are substantially modified due
to the response of the vacuum (the $\pm i\epsilon$ prescription is
designated for and is applicable only to the theories with the PT
vacua, indeed \cite{38,50}). Other regularization schemes are also
available, for example, such as analytical regularization used in
Ref. \cite{16} or the so-called Speer's regularization \cite{51}.
However, they should be compatible with the DT as emphasized
above. Anyway, not the regularization is important but the DT
itself. Just this theory provides an adequate mathematical
framework for the correct treatment of all the Green's functions
in QCD, and in quantum YM theory in particular (apparently, for
the first time the distribution nature of the Green's functions in
quantum field theory has been recognized and used in Ref.
\cite{52}).

The regularization of the NP IR singularities in QCD is determined
by the Laurent expansion (B4) at $n=4$ as follows:

\begin{equation}
(q^2)^{- 2 - k } = { 1 \over \epsilon} a(k)[\delta^4(q)]^{(k)} +
f.t. = { 1 \over \epsilon} \Bigr[ a(k)[\delta^4(q)]^{(k)} +
O(\epsilon) \Bigl], \quad \epsilon \rightarrow 0^+,
\end{equation}
where $a(k)$ is a finite constant depending only on $k$ and
$[\delta^4(q)]^{(k)}$ represents the $k$th derivative of the
$\delta$-function (see Eqs. (B4) and (B5)). We point out that
after introducing this expansion everywhere one can fix the number
of dimensions, i.e., put $D=n=4$ for QCD without any further
problems. Indeed there will be no other severe IR singularities
with respect to $\epsilon$ as it goes to zero, but those
explicitly shown in this expansion. Let us underline that, while
the initial expansion (6.1) is the Laurent expansion in the
inverse powers of the gluon momentum squared, the regularization
expansion (B7) is the Laurent expansion in powers of $\epsilon$.
This means that its regular part is as follows: $f.t. =(q^2)^{- 2
- k }_{-} + \epsilon (q^2)^{- 2 - k }_{-} \ln q^2 +
O(\epsilon^2)$, where for the unimportant here definition of the
functional $(q^2)^{- 2 - k }_{-}$ see Ref. \cite{12}. These terms,
however, play no any role in the IRMR program which has been
preliminary discussed in section VI. The regularization expansion
(B7) takes place only in four-dimensional QCD with Euclidean
signature. In other dimensions and/or Minkowski signature it is
much more complicated as pointed out above. As it follows from
this expansion any power-type NP IR singularity, including the
simplest one at $k=0$, scales as $1 /\epsilon$ as it goes to zero.
Just this plays a crucial role in the IR renormalization of the
theory within our approach. Evidently, such kind of the
dimensionally regularized expansion (B7) does not exist for the PT
IR singularity, which is as much singular as $(q^2)^{-1}$ only.

 In summary, first we have emphasized the
distribution nature of the NP IR singularities. Secondly, we have
explicitly shown how the DRM should be correctly implemented into
the DT. This makes it possible to put severe IR singularities
under firm mathematical control.

\section{The gluon SD equation and its general iteration solution}

Let us show explicitly that the general iteration solution (4.18)

\begin{equation}
D_{\mu\nu}(q; \Delta^2) = D^{INP}_{\mu\nu}(q; \Delta^2) +
D^{PT}_{\mu\nu}(q)
\end{equation}
satisfies the gluon SD equation (3.7)

\begin{equation}
D_{\mu\nu}(q) = D^0_{\mu\nu}(q) - T_{\mu\sigma}(q) \Bigl[\Pi(q^2;
D) + { \Delta^2(\lambda; D) \over q^2} \Bigr] D_{\sigma\nu}(q),
\end{equation}
indeed (let us recall that we replace $\Pi^s_1(q^2; D) \rightarrow
\Pi(q^2; D)$, for convenience). There is no doubt that our
solution for the full gluon propagator (C1), obtained at the
expense of remaining unknown its PT part, nevertheless, satisfies
the gluon SD equation (C2), since it has been obtained by the
direct iteration solution of this equation. To show this
explicitly by substituting it back into the gluon SD equation (C2)
is not a simple task, and this is to be done elsewhere. The
problem is that the decomposition of the full gluon propagator
into the INP and PT parts by regrouping the so-called mixed terms
in section IV was a well defined procedure (there was an exact
criterion how to distinguish between these two terms in a single
$D$). However, to do the same at the level of the gluon SD
equation itself, which is nonlinear in $D$, is not so obvious.

Fortunately, there exists a rather simple method as how to show
explicitly that the INP part of the full gluon propagator can be
completely decoupled from the rest of the gluon SD equation in the
$\epsilon \rightarrow 0^+$ limit. For this purpose, let us
consider it as a function of $\epsilon$ rather than as a function
of its momentum. There is no explicit integration over the gluon
momentum $q$ in the gluon SD equation (C2). As we already know, in
this case the INP part of the full gluon propagator vanishes as
$\epsilon$ goes to zero (see Eq. (6.18)). Then the full gluon
propagator becomes $D_{\mu\nu}(q) = D^{PT}_{\mu\nu}(q)$. At the
same time, the PT part $D^{PT}_{\mu\nu}(q)$ can be considered as
the IR renormalized from the very beginning, since it is free of
the NP IR singularities, by construction, i.e.,
$D^{PT}_{\mu\nu}(q) \equiv \bar{D}^{PT}_{\mu\nu}(q)$. On the other
hand, the invariant function $\Pi(q^2; D)$ has been obtained by
the corresponding subtraction procedure (see Eq. (3.3)). So it is
free of quadratic divergences parameterized in terms of the mass
gap squared (as mentioned above, it may have only logarithmic
divergences). In fact, this means that effectively it depends not
on $D$ but on $D^{PT}$, i.e., $\Pi(q^2; D) \equiv \Pi(q^2;
D^{PT})$. So Eq. (C2) becomes

\begin{equation}
D^{PT}_{\mu\nu}(q) = D^0_{\mu\nu}(q) - T_{\mu\sigma}(q)
\Bigl[\Pi(q^2; D^{PT}) + { \Delta^2(\lambda; D) \over q^2} \Bigr]
D^{PT}_{\sigma\nu}(q).
\end{equation}
In order to go to the $\epsilon \rightarrow 0^+$ limit in this
equation, it is necessary to express the mass gap
$\Delta^2(\lambda; D) = \Delta^2 \equiv \Delta^2(\lambda; \alpha,
\xi, g^2)$ through its IR renormalized counterpart with the help
of the definitions (6.7) and (6.15) which yield

\begin{equation}
\Delta^2 = Z^{-1} \Delta^2_R = \epsilon \times Z^{-1}
\bar{\Delta}^2_R, \quad \epsilon \rightarrow 0^+.
\end{equation}
Let us remind now that the UVMR constant $Z$ depends in general on
a set of parameters: $\xi$, $\lambda, \alpha_s(\lambda)$ in the
regime when both $\lambda$ and  $\alpha_s(\lambda)$ go to infinity
(see Eq. (7.2)). That is the gauge-fixing parameter $\xi$ is not
IR renormalized, i.e., $\xi = \bar{\xi}$ follows from the relation
$D^{PT}_{\mu\nu}(q) \equiv \bar{D}^{PT}_{\mu\nu}(q)$ mentioned
above. Thus $Z$ cannot depend on $\epsilon$ when it goes to zero.
This is in agreement with the theory of functions of complex
variable (see WSK theorem above), where it is considered as the
number, depending only on a sequence of points along which the
gluon momentum goes to zero. Apparently, in this case it is indeed
possible to distribute the dependence on $\epsilon$ between two
mass gaps $\Delta^2$ and $\Delta^2_R$, leaving thus relating
dimensionless parameter $Z^{-1}$ free of it. All this means that
after substituting of this relation into the previous gluon SD
equation, we can go to the $\epsilon \rightarrow 0^+$ limit. Then
the gluon SD equation finally becomes

\begin{equation}
D^{PT}_{\mu\nu}(q) = D^0_{\mu\nu}(q) - T_{\mu\sigma}(q) \Pi(q^2;
D^{PT}) D^{PT}_{\sigma\nu}(q).
\end{equation}
It is easy to check that its "solution" is the effective charge
(A6) again, where the mass gap is removed in accordance with the
relation (C4).

It is instructive to equivalently rewrite the previous equation as
follows:

\begin{equation}
D^{PT}(q) = D^0(q) + D^0(q)O(q^2; D^{PT})D^{PT}(q),
\end{equation}
where we introduce the following notation $O(q^2; D^{PT})= i q^2
\Pi(q^2; D^{PT})$, as well as omit the Dirac indices, for
convenience. By introducing it, we would like to underline that
this composition is always of the order $q^2$ at any $D^{PT}$ (let
us note that it will be of this order even at any $D$). The
nonlinear iteration solution of Eq. (C6) looks like

\begin{eqnarray}
D^{PT}(q) &=& D^0(q) + D^0(q)O(q^2; D^{PT})D^{PT}(q) \nonumber\\
&=& D^0(q) + D^0(q)O_1(q^2)D^0(q) +
D^0(q)O_0(q^2)D^0(q)O_1(q^2)D^0(q) + ...,
\end{eqnarray}
where $O_0(q^2)=O(q^2; D^0)$ and $O_1(q^2)= O(q^2; D^0 + D^{(1)})$
with $D^{(1)}(q)= D^0O_0(q^2)D^0$, and so on. Since the free gluon
propagator $D^0(q) \sim (1 / q^2)$, while all $O_n(q^2), \ n
=0,1,2,3,...$ are always of the order $q^2$, the iteration
solution (C7) will produce only the PT-type (i.e., $1/ q^2$) IR
singularities. This is in complete agreement with the IR structure
of the PT part of the full gluon propagator in Eq. (4.18).

Thus the INP part of the full gluon propagator is automatically
decoupled from the rest of the gluon SD equation as $\epsilon$
goes to zero at the final stage. Just in this sense should be
understood in general terms the solution of the gluon SD equation
within our approach, since it leaves the PT part of the solution
undetermined. To show this explicitly by treating the INP part of
the full gluon propagator as a function of its momentum and
substituting the corresponding decomposition back to the nonlinear
gluon SD equation is far more complicate case as mentioned above.


\begin{thebibliography}{}

\bibitem{1}
   W. Marciano, H. Pagels, Phys. Rep. C 36 (1978) 137.
\bibitem{2}
   M.E. Peskin, D.V. Schroeder, An Introduction to Quantum Field Theory \\
   (AW, Advanced Book Program, 1995).
\bibitem{3}
   Confinement, Duality, and Nonperturbative Aspects of QCD,
   edited by P. van Baal, \\
   NATO ASI Series B: Physics, vol. 368 (Plenum, New York, 1997).
\bibitem{4}
   Non-Perturbative QCD, Structure of the QCD vacuum, edited by
   K-I. Aoki, O. Miymura and T. Suzuki, Prog. Theor. Phys. Suppl.
   131 (1998) 1.
\bibitem{5}
   A. Jaffe, E. Witten, Yang-Mills Existence and Mass Gap, \\
   $http://www.claymath.org/prize-problems/, \
   http://www.arthurjaffe.com$ \ .
\bibitem{6}
   V.N. Gribov, J. Nyiri, Quantum Electrodynamics, (Cambridge
   University Press, 2001).
\bibitem{7}
   J.D. Bjorken, S.D. Drell, Relativistic Quantum Fields, (Mc
   Graw-Hill Book Company, 1978).
\bibitem{8}
   C. Itzykson, J.-B. Zuber, Quantum Field Theory, (Mc Graw-Hill
   Book Company, 1984).
\bibitem{9}
   T. Muta, Foundations of QCD, (Word Scientific, 1987).
\bibitem{10}
   V. Gogohia, hep-ph/0311061.
\bibitem{11}
   M. Baker, Ch. Lee, Phys. Rev. D 15 (1977) 2201.
\bibitem{12}
   I.M. Gel'fand, G.E. Shilov, Generalized Functions, (Academic
   Press, New York, 1968), Vol. I.
\bibitem{13}
   G. 't Hooft, M. Veltman, Nucl. Phys. B 44 (1972) 189.
\bibitem{14}
   V. Gogohia, Phys. Lett. B 584 (2004) 225.
\bibitem{15}
   V. Gogohia, Phys. Lett. B  618 (2005) 103.
\bibitem{16}
   H. Pagels, Phys. Rev. D 15 (1977) 2991.
\bibitem{17}
   L. Susskind, J. Kogut, Phys. Rep. C 23 (1976) 348.
\bibitem{18}
   H. Fritzsch, M. Gell-Mann, H. Leutwyler, Phys. Lett. B 47 (1973)
   365.
\bibitem{19}
   S. Weinberg, Phys. Rev. Lett. 31 (1973) 494.
\bibitem{20}
 H. Georgi, S. Glashow, Phys. Rev. Lett. 32 (1974) 438.
\bibitem{21}
   J.L. Gervais, A. Neveu, Phys. Rep. C 23 (1976) 240.
\bibitem{22}
   S. Mandelstam, Phys. Rev. D 20 (1979) 3223.
\bibitem{23}
   S.G. Matinyan, G.K. Savvidy, Nucl. Phys. B 134 (1978) 539.
\bibitem{24}
   V. Gogohia, Gy. Kluge, Phys. Rev. D 62 (2000) 076008.
\bibitem{25}
   M. A. Shifman, A. I. Vainshtein, V. I. Zakharov, Nucl. Phys. B 147
   (1979) 385, 448; \\
   V.A. Novikov, M. A. Shifman, A. I. Vainshtein, V. I. Zakharov, Nucl. Phys. B
   191 (1981) 301.
\bibitem{26}
   I. Halperin, A. Zhitnitsky, Nucl. Phys. B 539 (1999) 166.
\bibitem{27}
   E. Witten, Nucl. Phys. B 156 (1979) 269; \\
   G. Veneziano, Nucl. Phys. B 159 (1979) 213.
\bibitem{28}
   V. Gogohia, Phys. Lett. B 501 (2001) 60; \\
   V. Gogohia, H. Toki, Phys. Rev. D 61 (2000) 036006; ibid. 63
   (2001) 079901(E).
\bibitem{29}
   M. Gell-Mann, R.J. Oakes, B. Renner, Phys. Rev. 175 (1968) 2195.
\bibitem{30}
   V. Gogokhia, hep-ph/0606010, \ hep-th/0604095.
\bibitem{31}
   V. Gogohia, hep-ph/0508224; \\
   V. Gogohia, Gy. Kluge, Phys. Rev. D 62 (2000) 076008.
\bibitem{32}
   V. Gogohia, Gy. Kluge, M. Prisznyak, hep-ph/9509427.
\bibitem{33}
   M.A. Lavrentiev, B.V. Shabat, Methods of the theory of
   functions of complex variable, Russ. Ed., Moscow, Nauka, 1987.
\bibitem{34}
   K.B. Wilson, Phys. Rev. D 10 (1974) 2445.
\bibitem{35}
   M. Bander, Phys. Rep. 75 (1981) 205.
\bibitem{36}
   A.I. Alekseev, B.A. Arbuzov, Phys. Atom. Nucl. 61 (1998) 264.
\bibitem{37}
   K.D. Born et al., Phys. Lett. B 329 (1994) 325; \\
   V.M. Miller et al., Phys. Lett. B 335 (1994) 71.
\bibitem{38}
   V.N. Gribov, Gauge Theories and Quark Confinement (PHASIS,
   Moscow, 2002).
\bibitem{39}
   G. Preparata, Phys. Rev. D 7 (1973) 2973.
\bibitem{40}
   G. Pocsik, T. Torma, Acta Phys. Hun. 62 (1987) 101, 107; \\
   Z. Fodor, Acta Phys. Pol. B 19 (1988) 21.
\bibitem{41}
   G. 't Hooft, Nucl. Phys. B 75 (1974) 461.
\bibitem{42}
   I.V. Andreev, QCD and hard processes at high energies, Russ. Ed., Moscaw, Nauka, 1981.
\bibitem{43}
   F. Wilczek, Proc. Inter, Conf., QCD--20 Years Later, Aachen,
   June 9-13, 1992, v. 1.
\bibitem{44}
   D.V. Shirkov, hep-ph/0208082.
\bibitem{45}
   S. Coleman, E. Weinberg, Phys. Rev. D 7 (1973) 1888.
\bibitem{46}
   D.J. Gross, A. Neveu, Phys. Rev. D 10 (1974) 3235.
\bibitem{47}
   J.S. Ball, T.-W. Chiu, Phys. Rev. D 22 (1980) 2542, 2550.
\bibitem{48}
   G. Leibbrandt, Noncovariant Gauges (WS, Singapore, 1994).
\bibitem{49}
   A. Bassetto, G. Nardelli, R. Soldati, Yang-Mills Theories in
   Algebraic Non Covariant Gauges (WS, Singapore, 1991).
\bibitem{50}
   Y.L. Dokshitzer, D.E. Kharzeev, hep-ph/0404216.
\bibitem{51}
   E.R. Speer, Jour. Math. Phys. 15 (1974) 1.
\bibitem{52}
   N.N. Bogoliubov, D.V. Shirkov, Introduction to the Theory of
   Quantised Fields (Interscience Publisher, NY, 1959).
\end{thebibliography}
\end{document}